\documentclass[12pt]{article}

\usepackage[affil-it]{authblk}

\usepackage[in]{fullpage}

\usepackage[table]{xcolor}
\usepackage{hyperref}
\usepackage{times}
\usepackage{graphicx}
\usepackage{amssymb}
\usepackage{amsmath}
\allowdisplaybreaks
\usepackage[numbers]{natbib}
\usepackage[utf8]{inputenc}
\usepackage[T1]{fontenc}
\usepackage[USenglish]{babel}
\usepackage{url}
\usepackage{enumerate}
\usepackage{paralist}
\usepackage{here}
\usepackage{doi}
\usepackage{xspace}
\usepackage{paralist}
\usepackage{multirow}
\usepackage{siunitx}
\usepackage{tikz}
\usetikzlibrary{backgrounds,calc,positioning}

\newcommand{\RNum}[1]{\uppercase\expandafter{\romannumeral #1\relax}}

\xspaceaddexceptions{())]\}}
\newcommand{\POS}{POS\xspace}
\newcommand{\ARIMA}{ARIMA\xspace}
\newcommand{\RestA}{Restaurant A\xspace}
\newcommand{\RestB}{Restaurant B\xspace}
\newcommand{\restA}{restaurant A\xspace}
\newcommand{\restB}{restaurant B\xspace}
\newcommand{\Stan}{\textit{Stan}\xspace}
\newcommand{\RStan}{\textit{RStan}\xspace}
\newcommand{\R}{\textit{R}\xspace}
\newcommand{\maxf}[1]{{\cellcolor[gray]{0.8}} #1}
\newcommand{\transpose}[1]{\ensuremath{#1^{\mathsf{T}}}}
\newcommand{\starTranspose}[1]{\ensuremath{#1^{*\mathsf{T}}}}
\newcommand{\normalModel}{\textit{Normal}\xspace}
\newcommand{\negBinomModel}{\textit{NegBinom}\xspace}

\begin{document}

  \title{\bf A Bayesian Approach for Predicting Food and Beverage Sales in
  Staff Canteens and Restaurants \thanks{The research was by funded by the Austrian Research Promotion Agency (FFG)
  through the General Programme under the FFG grant number 867779.}}
  \author[1,2]{Konstantin Posch}
  \author[1]{Christian Truden
  \thanks{Corresponding author: Christian Truden, Email: \href{mailto:christian.truden@aau.at}{christian.truden@aau.at}}
   }
  \author[1]{Philipp Hungerl\"ander}
  \author[2]{J\"urgen Pilz}
  \affil[1]{Department of Mathematics, Alpen-Adria-Universität Klagenfurt, Austria}
  \affil[2]{Department of Statistics, Alpen-Adria-Universität Klagenfurt, Austria}

  \maketitle

\begin{abstract}
  Accurate demand forecasting is one of the key aspects for
  successfully managing restaurants and staff canteens.
  In particular, properly predicting future sales of menu items allows
  a precise ordering of food stock.
  From an environmental point of view, this ensures
  maintaining a low level of pre-consumer food waste,
  while from the managerial point of view, this is critical to
  guarantee the profitability of the restaurant.
  Hence, we are interested in predicting future values of the daily sold quantities of
  given menu items.
  The corresponding time series show multiple strong seasonalities,
  trend changes, data gaps, and outliers.
  We propose a forecasting approach that is solely based on
  the data retrieved from Point of Sales systems
  and allows for a straightforward human interpretation.
  Therefore, we propose two generalized additive models for predicting the future sales.
  In an extensive evaluation, we consider two data sets collected at a casual restaurant and a large staff canteen
  consisting of multiple time series, that cover a period of 20 months, respectively.
  We show that the proposed models fit the features of the considered restaurant data.
  Moreover, we compare the predictive performance of our method against the performance
  of other well-established forecasting approaches.
\end{abstract}
\noindent%
{\it Keywords: Sales Prediction; Time Series; Generalized Additive Models; Bayesian Methods; Shrinkage Priors}
\newpage

\section{Introduction}
\label{sec:intro}

The total turnover in the US restaurants sector is projected to reach $\$863\text{bn}$  in 2019,
contributing to  $4 \,\%$ of the gross domestic product \cite{NRA2019}.
Currently, the restaurant industry employs $15.3$ million people
at one million locations across the United States.
At the same time, the industry accounts for $11.4$ million tons of food waste
that constitutes an estimated value of $\$25\text{bn}$ going to waste \cite{ReFeed2019} every year.

Food waste is a critical socioeconomic problem considering that $11.8\,\%$  of the households
in the U.S. suffered from food insecurity in 2018 \cite{USDA2019}.
Furthermore, production, transportation, and disposal of unused food significantly impact
the environment  \cite{TONINI2018744, Hickey2014}.
On the other hand, food costs represent $28\,\%$ to $35\,\%$ of sales in restaurants.
Hence, reducing food waste can be critical to boost
profitability and to reduce the environmental impact of operating a restaurant.
Furthermore, also food service contractors
face similar challenges when providing meals at staff canteens, hospitals, etc.

Accurate demand forecasting is one of the key aspects for successfully managing
restaurants and, from an environmental point of view, maintaining a low level
of pre-consumer food waste.
In this work, we are interested in predicting future values of
the daily sold quantities of a given menu item.
Hence, we deal with time series that show multiple strong seasonalities,
trend changes, and outliers.
Traditionally, judgmental forecasting techniques, based on the managers experience,
are applied to estimate future demand at restaurants.
However, producing high quality forecasts is a time consuming and challenging task,
especially for inexperienced managers.
Hence, we aim for a data-driven approach that supports restaurant managers in their decision making.

Nowadays, most restaurants and canteens use
\textit{(electronic) Point of Sales} (\POS) systems
to keep track of all their sales.
Clearly, the resulting data inventories are valuable  sources
for various data science applications such as forecasting future sales numbers of menu items.
In general, many industries rely on \POS data to predict future demand.
However, especially in the retailing sector, it is hard to make beneficial use of
these estimates due to complicated supply chains and long lead times.
In contrast to that, the restaurant industry is characterized
by short lead times when ordering food stock and the absence of a complex supply chain.

Hence, we propose a forecasting approach that requires only \POS data,
i.e., weather information and special promotions are not considered.
Moreover, in order to ensure acceptance of the approach among restaurant managers,
the models ability for a straightforward human interpretation is critical.

The remainder of this paper is organized as follows.
First, in Section \ref{sec:problemDesc}, we provide a concise problem description.
In Section \ref{sec:relatedWork}, we briefly discuss related literature.
Then, in Section \ref{sec:methodology}, we propose our forecasting approach.
In Section \ref{sec:evaluation}, we evaluate the prediction quality of our approach
using real-world data of sales in a restaurant and a canteen.
Finally, Section \ref{sec:conclusion} concludes the paper and outlines future
research directions.

\section{Problem Description}
\label{sec:problemDesc}

This work is concerned with predicting the \text{daily} sold quantities
of given menu items.
This information, together with the recipes of the items,
is essential to effectively manage the ordering of food stock.
In our use-case, an \textit{(electronic) Point of Sales} (\POS) system is the
exclusive data source.
Typically, such systems create a new line item for each sold menu item together with a time stamp.
This way, large data inventories, consisting of thousands of line items, are created.
Each restaurant, let it be a small burger joint or a large company canteen,
usually has many different products on its menu.
While some products are offered all year long, others are only seasonally available.
Moreover, the demand for certain products differs over the year.

Each menu item is usually identified through an unique ID.
From the stored records, we can query the accumulated number of sales
of the different menu items for each day.
Hence, gathering the data required for our forecasting approach causes little to no
additional administrative burden.
In some cases, it is reasonable to do forecasting for product groups
rather than for individual menu items.
For example, a canteen may have a daily vegetarian option.
Each day a different dish, identified by its own ID, is served.

\subsection{Assumptions}
\label{sec:assumptions}

For now, we assume that the sales of individual menu items are independent
of each other. While this assumption is unlikely to be true for all types of
restaurants, e.g., some sides go along better with certain main dishes than others,
it is a reasonable simplifying assumption.

By design, a \POS system only documents sales.
The absence of recorded sales of a certain menu item on a given day may be due to several
reasons:
\begin{compactenum}[(i)]
  \item The restaurant was closed. In this case, there is no recorded sale of any item at all.
  \label{enu:closed}
  \item The item was not on the menu that day by choice or due to lack of stock.
  \label{enu:bychoice}
  \item Nobody wanted to order the product.
  \label{enu:notordered}
\end{compactenum}
Obviously, it makes a difference if a product was not sold due to
the fact that nobody wanted to order it (\ref{enu:notordered}),
or due to the reason that the product was not offered (\ref{enu:closed}, \ref{enu:bychoice}).
To distinguish between these to cases we treat an absence of sales caused by (\ref{enu:notordered})
as zero sales and the absence of sales caused by (\ref{enu:closed}) or (\ref{enu:bychoice})
as missing values. Setting the number of sales independently
of (\ref{enu:closed}), (\ref{enu:bychoice}), or (\ref{enu:notordered}) always
equal to zero can lead to inferior model fits, especially if the periods where the restaurant is closed,
or does not offer special products, are aperiodic, or do not show a common pattern.
For sure, this could be prevented by adding some information regarding why there are no sales to the model.
However, this would require to manage and store additional data other than the \POS data,
which we do not want to presuppose in this work.

We assume that the restaurant was opened for business on a given day
if there is at least one recorded sale of any menu item.
Usually, data available from a \POS system contains no records about when a product was added to and
when it was removed from the menu. Therefore, we assume that an item has been
removed from the menu if there is no recorded sale for $60$ or more days.

\section{Related Work}
\label{sec:relatedWork}

In this section,
we give a brief overview of related literature.
We focus on forecasting approaches used in the restaurant industry
as well as on the use of \POS data.

A general view of forecasting methods for restaurant sales and customer demand
is given in the recent review paper by \citet{LASEK2016}.
\citet{Ryu2003} compare several methods (moving average, multiple regression, exponential smoothing)
in order to estimate the daily dinner counts at a university dinning center.
In their case, a study shows that multiple regression gives the most accurate predictions.
Also, \citet{reynolds_econometric_2013}
use multiple regression to predict the annual sales volume of the
restaurant industry (and of certain subsegments).
\citet{forst_forecasting_1992} applies \ARIMA and exponential smoothing models
 to forecast the weekly sales (in USD) at a small campus restaurant.
\citet{hu_forecasting_2004} use \ARIMA methods to predict the
customer count at casino buffets.
\citet{cranage_comparison_1992} use exponential smoothing to predict the monthly sales (in USD) at a restaurant.
\citet{bujisic_effect_2017} analyze the effects of weather factors onto restaurant sales.
The paper also contains a review of forecasting approaches for restaurant sales
(with and without weather factors).
However, the authors point out that most work from the literature is concerned
with aggregated forecasts,
i.e., only predictions for product categories or weekly sales numbers are given, rather than forecasting sales of
individual menu items per day.

Moreover, \citet{Tanizaki2018} propose to use machine learning techniques in order
to forecast the daily number of customers at restaurants.
Their predictions are based on  \POS data that is enriched with external data, e.g., weather information and event data.
Similarly, \citet{Kaneko2016}
present a deep learning approach to construct a prediction model for the sales at supermarkets using \POS data.

So far, this short review shows that existing work in the literature is mostly
concerned with predicting aggregated numbers rather than the sale of individual items.
Moreover, it reveals that using \POS data is common practice in many industries.
However, our review exposes a gap in the existing literature as,
to the best of our knowledge, there exists no work concerned
with predicting sales of individual menu items at restaurants (based on \POS data).

Hence,
we extend our review onto prediction approaches dealing
with data showing similar features than ours.
The problem considered in this work is clearly a time series prediction problem.
Its most dominant characteristics are:  multiple strong seasonalities,
trend changes, data gaps, and outliers.
Thus, the application of classical time series models such as \ARIMA models or exponential
smoothing \cite{HOLT20045} is of limited use.
These models can only account for one seasonality and, moreover, require missing data points to be interpolated.
In principle, multiple seasonality patterns can be dealt with in exponential smoothing, see e.g., \citet{Gould2008}.
However, most available \ARIMA or exponential smoothing implementations, e.g., \cite{forecast2008, forecast2019}, do not include such features.

Therefore, \citet{Taylor2018}
present a \textit{decomposable time series model} that is similar to a
generalized additive model \cite{Hastie1987}.
It consists of three main components: trend, seasonality, and holidays.
The model is designed in such a way that it allows for a straightforward human interpretation
for each parameter allowing an analyst to adjust the model if necessary.
The approach is concerned with time series having
features such as multiple strong seasonalities, trend changes, outliers, and holiday effects.
As a motivating example they use a time series describing the number of events
that are created on Facebook every day.
However, this number is quite large in comparison to the daily sales in restaurants and canteens.
Thus, for Facebook's example it is valid to assign a normal distribution to the target variable.
However, this might be inferior in our use case, since small target values are quite common.
Note that the normal distribution $\mathcal{N}(\lambda,\lambda)$  has a  shape similar
to the Poisson distribution $Poisson(\lambda)$ for large $\lambda$.
In contrast to the normal distribution, the Poisson distribution is a classical distribution for count data.

\section{Methodology}
\label{sec:methodology}

In this section, we propose two Bayesian \textit{generalized additive models} (GAMs) for demand prediction.
The first one assumes a normally distributed target (also called response) $y$,
the second one assumes that the response $y$ follows a negative binomial distribution.
Both models include a trend function $g(t)$ and a seasonality function $s(t)$.
Sparsity inducing priors \cite{Park2008, CARLOS2010} are assigned to the parameters
of $g(t)$ and $s(t)$. This allows for a good model regularization, selection of
significant trend changes or influential seasonal effects, and, finally, enhanced model interpretability.

Considering that restaurants and canteens commonly offer many different menu items
and that new data is recorded on a daily basis, multiple models have to be
trained on a frequent basis.
For this reason, the training process should require only little computation time.
Additionally, the inference should take place automatically without the drawback
that a human analyst has to investigate the convergence of the procedure.
Therefore, we consider the application of full Bayesian inference as
inappropriate and focus on the comparatively easy to obtain mode of the posterior,
the so-called \textit{maximum a posteriori} (MAP) estimate, instead.

\subsection{Generalized Additive Models}

A generalized additive model writes as
$$g(\mathbb{E}(y))=\beta_0+\beta_1x_1+\ldots +\beta_kx_k+f_1(z_1)+\ldots +f_q(z_q),$$
where $y$ denotes the target variable corresponding to some exponential family
distribution, $x_1,\ldots , \allowbreak x_k,z_1,\ldots ,x_k \in \mathbb{R}$ denote the predictors
(also called covariates), $g$ denotes a link function (bijective and twice differentiable),
and $f_1,\ldots ,f_q$ denote some smooth functions. Typically, the functions $f_i$ are defined
as weighted sums of basis functions, i.e., $f_i=\sum_{k=1}^{K_i}\gamma_{ik}B_{ik}(z_i).$
Sparsity inducing priors can be used to perform variable selection within the set $\{x_1,\ldots ,x_k\}$
and, further, appropriate priors can be used to control the smoothness of the functions $f_1,\ldots ,f_q$.

\subsection{Trend Function}
\label{trend}

To model the trend of a time series, i.e., a series of data points ordered in time,
 we use a polynomial spline $g$ of degree $l$,
see \citet{Fahrmeir2013}.

A mapping $g:[a,b]\rightarrow \mathbb{R}$ is
called \textit{polynomial spline of degree} (order) $l\geq 0$ with knot points at $k_1<\ldots <k_m$, ($a<k_1,k_m<b$), if
\begin{compactenum}
  \item $g$ is a polynomial of degree $l$ on each of the intervals $[a,k_1)$, $[k_1,k_2)$,\ldots , $[k_m,b)$, and
  \item $g$ is $l-1$ times continuously differentiable provided that $l>0$.
\end{compactenum}
Let $S_l^\mathbf{k}$ denote the set of all $l$-th order splines with knots given by $k_1<\ldots <k_m$.
Equipped with the operations of adding two functions and taking real multiples $S_l^\mathbf{k}$ is a real vector space.
One can show that each element of $S_l^\mathbf{k}$ can be uniquely written as linear combination of the $d=l+m+1$ functions
\begin{align*}
  B_1(t)&=1,~B_2(t)=t,\ldots ,~B_{l+1}(t)=t^l, \\ B_{(l+2)}(t)&=(t-k_1)_+^l,\ldots ,~B_{(l+m+1)}(t)=(t-k_m)_+^l,
  \intertext{where}
  (t-k_j)_+^l &=
  \begin{cases}
    (t-k_j)^l, & t\geq k_j,\\
    0, & \text{else}.
  \end{cases}
\end{align*}
For this reason, the functions $B_1,\ldots ,B_d$ build a basis $B$ of the spline space $S_l^\mathbf{k}$.
$B$ is called the \textit{truncated power series basis} (TP-basis). The TP-basis allows for a simple interpretation of the trend model
\begin{align}
  \label{eq:TPbasis}
  g(t)&=\sum\limits_{j=1}^{d}\gamma_jB_j(t)\\
  &=\gamma_1+\gamma_2t+\ldots +\gamma_{l+1}t^l+\gamma_{l+2}(t-k_1)_+^l+\ldots +\gamma_{l+m+1}(t-k_{m})_+^l. \nonumber
\end{align}
The trend model $g$ consists of a global polynomial of degree $l$ which changes at each
knot $k_j$. The amount of change at a given knot point $k_j$ is determined by the absolute value
of the corresponding coefficient $\gamma_{l+j+1}$.
Thus, the knots of the polynomial spline are interpreted as (possible)
change points in the trend.

As a consequence, the assignment of a sparsity inducing prior to the trend coefficients
$\gamma_1,\gamma_2,\ldots , \allowbreak \gamma_d$ would allow for an identification of
truly significant trend changes.
This increases the interpretability of the model and also regularizes the trend function.

According to domain experts, for most menu items the trend (regarding the number of sales)
is quite constant over long periods, while significant changes appear only occasionally.
Hence, we decide to specify a knot point every $k$-th day between the first and
the last date with observations.
Assuming that the trend may change on a weekly basis, a possible value for $k$ could
be $k=7$. However, for restaurants that observe only slight changes once in a while also
larger values for $k$ are useful in order to reduce the size of the resulting model.
Consequently, we choose $k=30$ in our experimental evaluation, see Section \ref{sec:eval}.

\subsubsection{Prior for the Trend}
\label{priorTrend}
In this section, we propose two different priors for the trend model.
The first one is mainly inspired by the Bayesian Lasso \citep{Park2008}.
Here,  we assign independent Laplace priors to the coefficients of the trend model which
do not belong to the global polynomial.
In case that $y$ is assumed to be normally distributed with given variance $\sigma^2$ the priors read as
\begin{align}
  \label{eq:priorNormal}
  \gamma_{l+2}|\sigma^2,\ldots ,\gamma_{d}|\sigma^2\underset{i.i.d.}{\sim}Laplace\left(0,\frac{\sqrt{\sigma^2}}{\tau_1}\right).
\end{align}
The conditioning on $\sigma^2$  is important, because it guarantees a unimodal full posterior \cite{Park2008}.
Similarly, in case of a negative binomial response $y$ the priors are given by
\begin{align}
  \label{eq:priorNegBin}
  \gamma_{l+2},\ldots ,\gamma_{d}\underset{i.i.d.}{\sim}Laplace\left(0,\frac{1}{\tau_1}\right).
\end{align}
Assigning Laplace priors to coefficients $\gamma_{l+2},\ldots ,\gamma_{d}$ results in a sparse MAP estimate of these parameters.
The amount of sparsity is controlled by the tuning parameter $\tau_1$.
As a consequence, true change points can be automatically detected while wrongly proposed ones are ignored.
Moreover, as in Bayesian ridge regression we assign independent normal priors to the coefficients $\gamma_2,\ldots ,\gamma_{l+1}$:
\begin{align}
  \label{eq:normalPriorCoeff}
  \gamma_2,\ldots ,\gamma_{l+1}\underset{i.i.d.}{\sim}\mathcal{N}(0,\tau_3^2).
\end{align}
Thus, in the MAP estimate each of these coefficients is shrunken towards zero according
to its importance in a matter of regularization. Finally, the improper and
non-informative prior $p(\gamma_1)\propto 1$ is assigned to the intercept
of the trend function.

Besides the usage of Laplace and normal priors we also propose to assign the horseshoe prior,
see \citet{CARLOS2010}, to all the coefficients of the trend function except of the intercept.
In case of a Gaussian target variable the  prior reads as (see \cite{Makalic2016}):
\begin{align*}
  \gamma_j|\lambda_j,\tau,\sigma^2&\sim\mathcal{N}(0,\lambda_j^2\tau^2\sigma^2),\\
  % \label{eq:priorHorse1}\\
  \lambda_j&\sim C^{+}(0,1), \\
  % \label{eq:priorHorse2}\\
  \tau&\sim C^{+}(0,1),
  % \label{eq:priorHorse3}
\end{align*}
where $C^{+}$ denotes the half-Cauchy distribution and $j\in\{2,\ldots ,d\}$.
If a negative binomial response is assumed, $\sigma^2$ has to be removed from
above specification. The horseshoe prior is a shrinkage prior which enforces
a global scale $\tau$ on the one hand and, on the other hand, allows for
individual adaptions $\lambda_2,\ldots ,\lambda_d$ of the degree of overall shrinkage.

The possibility to use individual shrinkage strengths sometimes
is an advantage of the horseshoe prior over the Bayesian Lasso, where the shrinkage
effect is uniform across all coefficients. We notice that for some menu items the
trend of the corresponding time series changes drastically within a short period of time.
In this case uniform shrinkage either leads to an over-shrinkage of heavy changes,
or to an under-shrinkage of insignificant changes.
However, it turned out that a model using the horseshoe prior requires a very good
initialization of the respective optimization routine, in order for the procedure
to converge  towards a useful MAP-estimate.
Since we could not identify an automatic approach to find such an initialization,
we recommend to use the Bayesian Lasso prior as standard approach, and let
an expert apply the horseshoe prior if necessary.

Taking into account that the application of the horseshoe prior leads to problems
in terms of automation we have not considered other similarly sophisticated priors
\cite{Veronika2018, Bhattacharya2015} for now.
 However, this is subject to future work.

\subsubsection{Dates}
Obviously, the trend model requires the time $t$ to be given as numeric values.
Hence, we use a function $\delta:\{dates\}\rightarrow \mathbb{R}$ that maps all
possible dates (time points, days) to real numbers. We assume that observations
are available at the dates  $t_1<\ldots <t_n$.
Further, let  $N \geq n$ denote the number of days that lie in the
interval $[t_1,t_n]$.
Then, $\delta$ maps $t_1$ to zero and each other date $t$ to the number of days
that $t_1$ differs from $t$ divided by $N$.
Finally, $\delta(t_1),\ldots ,\delta(t_n)$ are elements of the interval $[0,1)$.

\subsection{Seasonality Function}
\label{seasonality}
The seasonality function $s(t)$ is used to model periodic changes of the daily
sold quantities $y_t$. In order to take account of the seasonal effects
\textit{day of week}, \textit{day of month}, and \textit{month of year} diverse dummy variables
(taking only the values $0$ and $1$ in order indicate the presence of a given season) are introduced.
Let $\mathbf{x}=\transpose{(x_1,\ldots ,x_s)}$ denote the vector consisting of all dummy variables
corresponding to the considered seasonal effects. The seasonality function is then given by
\begin{align*}
  % \label{eq:seaonalityFct}
  s(t)=\beta_1x_1(t)+\ldots +\beta_sx_s(t)=\transpose{\boldsymbol\beta}\mathbf{x}(t).
\end{align*}

\subsubsection{Prior for the Seasonality}
\label{priorSeason}

Inspired by the Bayesian Lasso \cite{Park2008}, we assign independent Laplace priors to the coefficients of the seasonality model.
For a normally distributed target $y$, the priors are specified as
\begin{align*}
  % \label{eq:priorSeasonNormal}
  \beta_1|\sigma^2,\ldots ,\beta_s|\sigma^2\underset{i.i.d.}{\sim}Laplace\left(0,\frac{\sqrt{\sigma^2}}{\tau_2}\right),
\end{align*}
and for a negative binomial target the priors read as
\begin{align*}
  % \label{eq:priorSeasonNegBin}
  \beta_1,\ldots ,\beta_s\underset{i.i.d.}{\sim}Laplace\left(0,\frac{1}{\tau_2}\right).
\end{align*}
The sparsity of the parameter's MAP estimate induced by the Laplace priors enables an
automatic selection of influential seasonal effects. We do not consider the usage
of the horseshoe prior \citep{CARLOS2010} for the seasonal model.
As already mentioned in Section \ref{priorTrend}, this prior results in a
more complicated model optimization.
Since we discovered empirically that the Bayesian Lasso prior
performs quite well for the seasonal model it appears unnecessary to accept the additional complexity.

\subsubsection{Deciding the Granularity of the Seasonality Model}

Let the number of days with observations be denoted by $n$.
Clearly, the size of the number $n$ influences the usefulness of modeling diverse seasonal effects. Hence, we apply the following strategy:
\begin{compactitem}
  \item $n<30$: model only the effect of \textit{day of week},
  \item $30\leq n<120$: additionally model the effect of \textit{month of year},
  \item $n>=120$: model all seasonal effects (\textit{day of week}, \textit{day of month}, and \textit{month of year}).
\end{compactitem}

\subsection{Prediction Models}

As already mentioned at the beginning of Section \ref{sec:methodology}, we propose two Bayesian GAMs
for predicting the daily sold quantities of certain menu items in restaurants and canteens.
Both models include a trend function $g(\delta(t))$
(see Section \ref{trend}) and a seasonality function $s(t)$ (see Section \ref{seasonality}).

\subsubsection{Normal Model}

In the so-called normal model, the daily sold quantity at time $t$, denoted by $y_t$, is modeled as
\begin{align}
  \label{eq:normalModel}
  y_t = g(\delta(t))+s(t)+\varepsilon_t,
\end{align}
where  $\varepsilon_t$ denote noise terms that are assumed to be i.i.d. according to
$\mathcal{N}(0,\sigma^2)$. Additionally, the non-informative and scale-invariant
prior $p(\sigma^2)\propto\frac{1}{\sigma^2}$ is assigned to the unknown error variance.
Suggestions for priors used to regularize $g(\delta(t))$ and $s(t)$ are described in
Section \ref{priorTrend} and Section \ref{priorSeason}. Assume that we have observations at the time points $t_1<\ldots <t_n$,
then Equation   \eqref{eq:normalModel}
translates to
\begin{align}
  \label{eq:normalModelSimply}
  \mathbf{y}|\boldsymbol\beta,\boldsymbol\gamma,\sigma^2&~~
  \sim~~\mathcal{N}(\mathbf{X}\boldsymbol\beta+\mathbf{Z}\boldsymbol\gamma,\sigma^2\mathbf{I}),
\end{align}
with
\begin{align}
  \label{eq:matrixX}
  \mathbf{X}=
  \begin{pmatrix}
    x_1(t_1) & x_2(t_1) & \cdots & x_s(t_1)\\
    \vdots &\vdots& & \vdots\\
    x_1(t_n) & x_2(t_n) & \cdots & x_s(t_n)
    \end{pmatrix},
  \end{align}
  according to Section \ref{seasonality} and
  \begin{align}
    \label{eq:matrixZ}
    \mathbf{Z}=
    \begin{pmatrix}
      1 & \delta(t_1) & \cdots & \delta(t_1)^l & (\delta(t_1)-\delta(k_1))_+^l & \cdots (\delta(t_1)-\delta(k_m))_+^l\\
      \vdots&   \vdots  &        &  \vdots     &     \vdots    & \vdots\\
      1 & \delta(t_n) & \cdots & \delta(t_n)^l & (\delta(t_n)-\delta(k_1))_+^l & \cdots (\delta(t_n)-\delta(k_m))_+^l
    \end{pmatrix}
    ,
  \end{align}
  according to Section \ref{trend}.

  \subsubsection{Negative Binomial Model}
  \label{sec:negBinomMod}

  As can be seen in Equation    \eqref{eq:normalModelSimply}, the assumption of additive and normally
  distributed noise directly implies  that the target $y$ also follows a normal distribution.
  In our use-case this assumption cannot be satisfied since the target is a non-negative integer.
  Indeed, this can even lead to predicting negative values,
  in case that the trend is a monotonically decreasing function.
  More appropriate distributions for the target variable are the
  Poisson distribution and the negative binomial distribution,
  which are both well-established distributions for regression with count data.

  Nevertheless, it should be mentioned, that, for large $\lambda$,
  the normal distribution $\mathcal{N}(\lambda,\lambda)$ is a good approximation
  of the Poisson distribution $Poisson(\lambda)$ due to the central limit theorem.
  However, the mean number of daily sold quantities can be quite small for some menu items in restaurants and staff canteens.
  Thus, we decide to model the target $y$, additionally to the normal
  distribution, also with a negative binomial distribution.
  This distribution can be considered as an extension of the Poisson distribution
  that also allows for over-dispersion. The mean and the variance of $y$ are given by
  \begin{align*}
    \mathbb{E}(y)&=\mu,\\
    \operatorname{Var}(y)&=\mu+\mu^2/\phi,
  \end{align*}
  where $\mu,\phi\in\mathbb{R}^{+}$ denote the parameters of the distribution.

  Although, the negative binomial distribution does not allow for under-dispersion, we do not consider this
  limitation as a major drawback.
  Under-dispersed data is rather unlikely in practical applications \cite{Hilbe2017}.
  Domain experts have confirmed this statement with regards to the application
  considered in this work. In other applications, where under-dispersion is common,
  the Conway–Maxwell–Poisson-distribution \cite{Conway1962} is a notable alternative to
  the negative binomial distribution. This distribution allows for both under-dispersed and over-dispersed data,
  but at the price of significantly increased model complexity.

  In the so-called negative binomial model the quantity $y_t$ sold on day $t$ is modeled as
  \begin{align}
    \label{eq:quantityNegBin}
    y_t|\boldsymbol\beta,\boldsymbol\gamma,a\sim NegBinom\left(\mu_t = \operatorname{exp}(g(\delta(t))+s(t)) , \phi =\frac{1}{a^2}\right),
  \end{align}
  with $a>0$.
  Using the exponential function $\operatorname{exp}()$ as response function
  ensures that the expected demand $\mu_t$ always stays positive, no matter
  how the estimates of $g(\delta(t))$ and $s(t)$ look like.
  Another side effect of this response functions is that the seasonal component
  acts in a multiplicative way on the trend, $\mu_t = \operatorname{exp}(g(\delta(t)))\operatorname{exp}(s(t))$.
  As recommended by \citet{Gelman2019} we assign a standard half-normal prior $\mathcal{N}^+(0,1)$ to $a = \frac{1}{\sqrt{\phi}}$:
  \begin{align}
    \label{eq:priorHalfNormal}
    a\sim \mathcal{N}^+(0,1).
  \end{align}
  Note that by assigning the prior $\mathcal{N}^+(0,1)$ directly to $\phi$,
  most of the prior mass would be on models with a large amount of over-dispersion.
  In case of limited over-dispersion, this can lead to a conflict between prior and data.

  Assuming that observations are available at the time points $t_1<\ldots <t_n$ model     \eqref{eq:quantityNegBin}
  translates to
  \begin{align}
    \label{eq:quantityNegBinSimple}
    \mathbf{y}|\boldsymbol\beta,\boldsymbol\gamma,a\sim\prod\limits_{i=1}^{n}\operatorname{NegBinom}\left(\operatorname{exp}\left\{\mathbf{X}[i,:]
    \boldsymbol\beta+\mathbf{Z}[i,:]\boldsymbol\gamma\right\},\frac{1}{a^2}\right),
  \end{align}
  where $\mathbf{y}=\transpose{(y_{t_1},\ldots ,y_{t_n})}$, $\mathbf{X}[i,:]$
  denotes the $i$-th row of matrix $\mathbf{X}$ defined via \eqref{eq:matrixX}, and $\mathbf{Z}[i,:]$
  denotes the $i$-th row of matrix $\mathbf{Z}$ defined via \eqref{eq:matrixZ}.

  \subsection{Choice of the Tuning Parameters}
  \label{choiceTuning}

  The priors proposed for the seasonality function $s(t)$ (see Section \ref{priorSeason})
  depend on a hyperparameter $\tau_2$. Further, in case that the priors      \eqref{eq:priorNormal} -- \eqref{eq:normalPriorCoeff}
  are used for the trend $g(\delta(t))$, they depend on hyperparameters denoted by
  $\tau_1$ and $\tau_3$. These parameters can be considered as tuning parameters
  that control the amount of sparsity/regularization in the estimates of $g(\delta(t))$ and $s(t)$.
  Thus, a good specification of $\tau_1,\tau_2,$ and $\tau_3$ is essential for the
  model performance. Bad choices could either allow for too much flexibility within
  a given model (inclusion of too many seasonal effects, a trend function that nearly
  interpolates the data, etc.) and, thus, result in overfitting, or do not allow
  for enough flexibility such that some significant effects are ignored.

  In principle, there exist three approaches to specify the hyperparameters $\tau_1,\tau_2,\tau_3$:
  \begin{compactitem}
    \item The first one is to determine standard settings that perform quite well on average.
    \item Another possibility is to determine useful settings via cross-validation.
    While, in general, the prediction quality of the resulting model is better than for
    the standard settings, the process itself is computationally expensive.
    \item The third option would be to specify hyper-priors for the tuning parameters.
  \end{compactitem}
  In this work, we focus on the first two approaches and defer the last one to future work.

  \subsubsection{Standard Settings}
  \label{standard}

  Certainly, good specifications of the tuning parameters differ from data set to
  data set, i.e., in the context of this work from menu item to menu item.
  Nevertheless, it is useful to have a fixed specification of them which works quite well on average.
  For this purpose, we have tested diverse possible specifications of these
  parameters with several data sets of representative menu items.
  The following specifications led to the best results:
  \begin{align}
    \label{eq:standardSettings}
    \tau_1 = 5,\qquad
    \tau_2 = 6, \qquad
    \tau_3 =\begin{cases}
    0.001, & \text{if } n<120,\\
    0.01, & \text{if } 120\leq n<350,\\
    0.5, & \text{else}.
  \end{cases}
\end{align}

\subsubsection{Cross-Validation}
\label{cv}

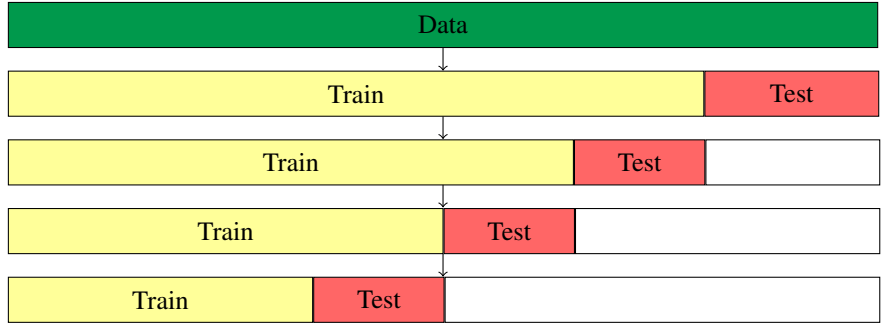
\begin{figure}
  \centering
  % !TeX spellcheck = en_US
\definecolor{fgreen}{RGB}{0,153,76}
\definecolor{yels}{RGB}{255,255,153}
\definecolor{lob}{RGB}{255,102,102}

\begin{tikzpicture}[
  block/.style={
  draw,
  fill=white,
  rectangle,
  minimum height=0.6cm,
  font=\footnotesize}]

  \def\mywidth{0.7*\textwidth}

  \node[block,minimum width=\mywidth, fill=fgreen](one) at (0,0) {Data} ;

  \node[block, below right=0.3cm and 0.0cm of one.south west,minimum width=0.8*\mywidth, fill=yels](two){Train};
  \node[block, right=0.0cm of two.east ,minimum width=0.2*\mywidth, fill=lob](twotwo){Test};

  \node[block, below right=0.3cm and 0.0cm of two.south west,minimum width=0.65*\mywidth, fill=yels](three){Train};
  \node[block, right=0.0cm of three.east ,minimum width=0.15*\mywidth, fill=lob](threetwo){Test};
  \node[block, right=0.0cm of threetwo.east ,minimum width=0.2*\mywidth](twothree){};

  \node[block, below right=0.3cm and 0.0cm of three.south west,minimum width=0.5*\mywidth, fill=yels](four){Train};
  \node[block, right=0.0cm of four.east ,minimum width=0.15*\mywidth, fill=lob](fourtwo){Test};
  \node[block, right=0.0cm of fourtwo.east ,minimum width=0.35*\mywidth](fourthree){};

  \node[block, below right=0.3cm and 0.0cm of four.south west,minimum width=0.35*\mywidth, fill=yels](five){Train};
  \node[block, right=0.0cm of five.east ,minimum width=0.15*\mywidth, fill=lob](fivetwo){Test};
  \node[block, right=0.0cm of fivetwo.east ,minimum width=0.5*\mywidth](fourthree){};

  \draw [->] (one.south) to  node[auto] {} ++ (0,-0.3cm);
  \draw [->] (one.south)++ (0,-0.91cm) to  node[auto] {} ++ (0,-0.3cm);
  \draw [->] (one.south)++ (0,-1.82cm) to  node[auto] {} ++ (0,-0.3cm);
  \draw [->] (one.south)++ (0,-2.74cm) to  node[auto] {} ++ (0,-0.3cm);

\end{tikzpicture}
  \caption{Illustration of the cross-validation approach. The data is assumed to be ordered temporarily.}
  \label{fig:cv}
\end{figure}

Cross-validation allows for an automatic identification of good and individual tuning parameter
specifications for each data set.
Applying cross-validation on a given data set means that the data set is
partitioned into multiple train/test splits.
On each of these splits the considered model is trained and tested with different
specifications of its tuning parameters.
Finally, the specification which on average goes along with the best predictions
(according to some pre-defined quality measure) is selected.
To take account of temporal dependencies, we consider an expanding window
approach for defining the train/test splits.
The approach is outlined in Figure \ref{fig:cv}. Looking at the figure the
amount of training data is gradually reduced
while the amount of test data is kept constant.

\subsubsection{Overall Strategy}
\label{overallStrategy}

% As already mentioned at the beginning of Section \ref{choiceTuning}, specifying
% the hyperparameters $\tau_1,\tau_2,\tau_3$ via cross-validation instead of using
% standard settings generally leads to a better model performance.
% However, the price to pay is an increased computational effort.
% Considering that restaurants and canteens offer many different menu items and,
% thus, require also many prediction models, a fast computation time is essential.
% For this reason,

We recommend to use the standard settings proposed in
Section \ref{standard}, and perform a cross-validation only for
the most important menu items if the model performance is insufficient.
The restriction to the standard settings
should not be considered too critical, since we discovered empirically that they perform quite well.

In case that cross-validation is applied  to specify $\tau_1,\tau_2,\tau_3$,
we perform the process step by step for each single tuning parameter.
Optimizing all three tuning parameters at once would require the training of substantially
more models and is therefore considered as computationally too expensive. Within the step-wise
procedure at first the parameter $\tau_3$ (responsible for the global trend) is optimized,
then the parameter $\tau_1$ (responsible for trend changes) and, finally,
the parameter $\tau_2$ (responsible for seasonal effects).
Parameters that are not currently optimized are either fixed with the
standard settings provided  in \eqref{eq:standardSettings}, or, if available,
with the already optimized values.

Finally, it should be mentioned that the quality criterion used inside the cross-validation is
the \textit{mean absolute deviation}  (MAD) between the predicted values and the
corresponding true ones.

\subsection{Prediction \& Prediction Uncertainty}
\label{uncertainty}

Assume that $y_t$ has to be predicted for $t\in\{t_{1}^*,\ldots ,t_{p}^*\}$
with $t_1^*<\ldots <t_p^*$ and, further, assume that observations are available
at the time points $t_1<\ldots <t_n$ with $t_1\leq t_1^*$. Let $\widehat{\boldsymbol\beta},\widehat{\boldsymbol\gamma},\widehat{\sigma}$,
and $\widehat{a}$  denote MAP estimates corresponding either to the normal
model    \eqref{eq:normalModelSimply}
or to the negative binomial model     \eqref{eq:quantityNegBinSimple}.
Additionally, treat the estimates as if they were the true values. Then the expected value
\begin{align}
  % \label{eq:expectedValue1}
  \boldsymbol{\mu}^*&=\mathbb{E}(\mathbf{y}^*=\transpose{(y_{t_1^*},\ldots
  ,y_{t_p^*})}|\widehat{\boldsymbol\beta},\widehat{\boldsymbol\gamma})
  \nonumber,\\
  \label{eq:expectedValue2}
  &=
  \begin{cases}
    \mathbf{X}^*\widehat{\boldsymbol\beta}+\mathbf{Z}^*\widehat{\boldsymbol\gamma}, &\text{normal model},\\
    \operatorname{exp}(\mathbf{X}^*\widehat{\boldsymbol\beta}+\mathbf{Z}^*\widehat{\boldsymbol\gamma}),&
    \text{negative binomial model},
  \end{cases}
\end{align}
can be used for prediction. Note that the exponential function in
Equation    \eqref{eq:expectedValue2}
is taken component-wise and, moreover, that $\mathbf{X}^*$
and $\mathbf{Z}^*$ are defined  analogously to  $\mathbf{X}$ and $\mathbf{Z}$
(Equations \eqref{eq:matrixX} and \eqref{eq:matrixZ}), by replacing $t_1,\ldots ,t_n$ with $t_1^*,\ldots ,t_p^*$.

Besides predicting a posteriori reasonable values for $\mathbf{y}^*$,
providing some uncertainty information regarding the forecasts is essential for
planning food stock orderings.
Especially, intervals that contain the $y_{t_i^*}$ $(i=1,\ldots ,p)$ with a
pre-defined probability are of particular interest.
Then, dependent on the importance of the availability of a given menu item,
the  manager can decide to go along with the prediction
$\mu_i^*=\mathbb{E}(y_{t_i^*}|\widehat{\boldsymbol\beta},\widehat{\boldsymbol\gamma})$,
or adjust her or his orders closer to the upper or lower interval boundary. Recall (Section \ref{trend}
and Section \ref{priorTrend}) that the trend is assumed to be constant most of
the times but changes occasionally.
To model this assumption, the trend
function $g(\delta(t))$ includes a knot every $k$-th day between the
first and the last date with observations. Additionally, sparsity inducing
priors are assigned to the coefficients $\gamma_{l+2},\ldots ,\gamma_{d}$
which represent the trend changes at the knot points $k_1,\ldots ,k_m$.
The restriction of merely specifying knots in the observed time period,
implies that the trend stays constant in unobserved periods, i.e., in the future.
However, in case that prediction
uncertainty information is required, possible future trend changes must be considered.
Hence, we extend the trend function to
\begin{align}
  \label{eq:trendFunctUncerntainty}
  g^{*}(\delta(t)) =~ &\gamma_1+\gamma_2\delta(t)+\ldots +\gamma_{l+1}\delta(t)^l+\gamma_{l+2}(\delta(t)-\delta(k_1))_+^l+\\
  &\ldots +\gamma_d(\delta(t)-\delta(k_m))_+^l+\gamma_{d+1}(\delta(t)-\delta(k_{m+1}))_+^l+ \nonumber \\
  & \ldots +\gamma_{d+r}(\delta(t)-\delta(k_{m+r}))_+^l, \nonumber
\end{align}
where $k_{m+1},\ldots ,k_{m+r}$ denote additionally introduced knot points.
The additional knots extend the approach of specifying a knot point every $k$-th
day within the observed period to the time interval  $[t_1,t_p^*]\cup [t_1,t_n]$.
In case that $t_p^*-k_m<k$ the expansion $g^*$ is equal to the original trend
function $g$. Additionally, it is assumed that the coefficients
$\gamma_{d+1},\ldots ,\gamma_{d+r}$ are i.i.d. zero mean Laplace distributed:
\begin{align}
  \label{eq:coeffsZeroMeanLaplace}
  \gamma_{d+1},\ldots ,\gamma_{d+r}\underset{i.i.d.}{\sim} Laplace(0,b).
\end{align}
Assuming that the  scale of future trend changes is determined by the scale
of the historically observed changes $\widehat{\gamma}_{l+2},\ldots ,\widehat{\gamma}_{d}$
the parameter $b$ is estimated as
\begin{align}
  \label{eq:paramBEstim}
  \widehat{b}=\frac{1}{m}\sum\limits_{i = 1}^{m}|\widehat{\gamma}_{l+i+1}|.
\end{align}
Note that given i.i.d. samples $x_1,\ldots ,x_N$ from a zero mean Laplace distribution
with scale parameter $b$, the mean of the absolute values of the $x_i$ is the
maximum likelihood estimator of  $b$.
Since the estimates $\widehat{\boldsymbol\beta},\widehat{\boldsymbol\gamma},\widehat{\sigma}$,
and $\widehat{a}$ are treated as if they were the true values, $(1-\alpha)$
prediction intervals for the $y_{t_i^*}$ $(i=1,\ldots ,p)$ can be computed as follows:
\begin{compactenum}
  \item Expand the trend function to $g^*$, see Equation \eqref{eq:trendFunctUncerntainty}.
  \item By replacing $t_1,\ldots ,t_n$ with $t_1^*,\ldots ,t_p^*$:
  \begin{compactenum}
    \item Determine the model matrix $\mathbf{X}^*$ analogous to the computation of $\mathbf{X}$ (Equation \eqref{eq:matrixX}).
    \item Determine the model matrix $\mathbf{Z}^{**}$ analogous to the computation of $\mathbf{Z}$.
    Note that that now $g^*$ is used to model the trend and not $g$.
  \end{compactenum}
  \item If $g^*$ differs from $g$, sample from $\boldsymbol{\gamma}^*=\transpose{(\gamma_{d+1},\ldots
  ,\gamma_{d+r})}$ according to the Equations   \eqref{eq:coeffsZeroMeanLaplace} and   \eqref{eq:paramBEstim}.
  Let the samples be denoted by $\tilde{\boldsymbol{\gamma}}^*=\transpose{(\tilde{\gamma}_{d+1},\ldots
  ,\tilde{\gamma}_{d+r})}$.
  \item Draw a sample from the conditional distribution of $\mathbf{y}^*=\transpose{(y_{t_1^*},
  \ldots ,y_{t_p^*})}$ conditioned on $\tilde{\boldsymbol{\gamma}}^*$. The distribution is given by:
  \begin{align*}
    \label{eq:sampleDist}
    \begin{small}
      \mathbf{y}^*|\tilde{\boldsymbol{\gamma}}^*\sim
      \begin{cases}
        \mathcal{N}(\mathbf{X}^*\widehat{\boldsymbol\beta}+\mathbf{Z}^{**}
        \transpose{
        (\transpose{\widehat{\boldsymbol\gamma}},\starTranspose{ \tilde{\boldsymbol{\gamma}} }) },\widehat{\sigma}^2\mathbf{I}), &
        \text{normal},\\
        % \text{normal model},\\
        \prod\limits_{i=1}^{p}\operatorname{NegBinom}\left(\operatorname{exp}
        \left\{\mathbf{X}^*[i,:]\widehat{\boldsymbol\beta}+
        \mathbf{Z}^{**}[i,:]
        \transpose{(\transpose{\widehat{\boldsymbol\gamma}},
        \starTranspose{\tilde{\boldsymbol{\gamma}}})  }\right\},\frac{1}{\widehat{a}^2}\right),& \text{neg. bin.}
        % \text{negative binomial model.}
      \end{cases}
    \end{small}
  \end{align*}
  In case that $g^*$ equals $g$, the vector $\boldsymbol{\gamma}^*$ is empty and can be ignored.
  \item Repeat step 3 and step 4 a pre-defined number of times.
  \item Compute the $\alpha/2$ and $(1-\alpha/2)$ quantiles of the
  samples drawn from $\mathbf{y}^*|\tilde{\boldsymbol{\gamma}}^*$
  component-wise in order to obtain the desired prediction intervals.
\end{compactenum}

It should be mentioned that the assumptions taken for the uncertainty estimation
are rather strong and, therefore, one cannot expect the prediction intervals to have
exact coverage. Therefore, the intervals should
be considered as an indicator for the level of uncertainty.

At first, the restriction to MAP
estimates implies that uncertainty is underestimated. However, as already
stated at the beginning of Section \ref{sec:methodology}, we consider the
application of full Bayesian inference as inappropriate due to the additional complexity.
Moreover, it is restrictive to assume that future trend changes are independent
and of the same average magnitude as historic changes.
Nevertheless, considering
that restaurants and canteens commonly plan their stock orders $7$-$14$ days in advance,
the number of possible future trend changes that have to be taken account of is very limited.
For this reason, using overly sophisticated approaches to model future trend changes is not required.

Besides using $\alpha\%$ prediction intervals to measure the uncertainty of
future predictions they can also be used to validate if the model fits the observed data.
In a first step, the intervals are computed for the time period with observations.
Then we check if the fraction of days for which the target variable $y$
does not lie within the corresponding intervals is significantly larger than $(1-\alpha)\%$.
If this is the case the model does not fit the data well and must be adjusted.

\section{Performance Evaluation}
\label{sec:evaluation}

In this section, the prediction quality of our approach is evaluated.
First,  Subsection \ref{ssec:data} provides a brief description of the test data used.
Then, in Subsection \ref{ssec:implementation} we briefly explain how the proposed
models have been implemented.
In Subsection \ref{ssec:eval1}, we evaluate how the models fit the
features of the considered restaurant data.
Finally, in Subsection \ref{ssec:eval2}, we compare our prediction models  against other
promising approaches from the literature.

Throughout this section the degree $l$ of the polynomial spline used to model the
trend is specified as $l=1$. It turned out  that higher degrees often cause too
strong trend decreases or increases during prediction.
Moreover, the trend function is designed to have a knot point every $30$-th day.

\subsection{Testing Data}
\label{ssec:data}
In this study, \POS data from two different restaurants, that was provided by a
partnering restaurant consulting firm, is used.
\RestA is a rather casual place located in Vienna, Austria, offering a traditional Austrian menu.
\RestB is a large staff canteen in the Netherlands, operated by
a major food services company.
For both locations, the data covers a time span of about $20$ months.
Moreover, the most representative menu items are selected, i.e.,
accounting for most sales. For \restA the menu items are categorized.
An overview of the time series data considered per category is given in
Table \ref{tab:itemsRestA}.
Additionally, this table provides some
descriptive statistics of the time series data. In particular, \textit{mean}, \textit{standard deviation} (sd),
\textit{minimum} (min), \textit{maximum} (max), and the \textit{number of missing values} (miss) are presented.
These statistics are computed for the time series with the highest (indicated with H)
and the time series with the lowest (indicated with L) sales revenue, respectively per category.
The temporal order of the data is ignored for
the calculation of the mean and the standard deviation.
The idea of the statistics is simply to give the reader a feeling in which range the
individual series take values etc.
For \restB a categorization is not possible, since
we only obtained data for which the product names and categories have been masked.
Similar to \restA we provide an overview of the corresponding time series data
in Table \ref{tab:itemsRestB}.
We notice that \restB shows a high number of missing values since
it is closed on weekends (compare to Section \ref{sec:assumptions}).

\setlength\tabcolsep{3pt}

\begin{table}[!ht]
  \centering
  \begin{scriptsize}
    % !TeX spellcheck = en_US

\begin{tabular}{l  S S S S S S S SSSSS[table-format=2]  }
  \hline
  {Category}   &  {\# time series} &{meanH} & {meanL} & {sdH} & {sdL} & {minH} & {minL} & {maxH} & {maxL} &{missH} & {missL}\\
  \hline
  Starters   				&5     &10.008  &3.265    &6.927  &3.079  &0&0&47&17 &25&25\\
  Side dishes				&7     &9.717   &1.786    &4.865  &1.627  &0&0&29&9 &25&25\\
  Main dishes               &21    &14.980  &1.868    &7.294  &1.671  &0&0&50&9 &25&25\\
  Desserts                  &7     &7.151   &0.875    &5.980  &1.373  &0&0&41&18 &25&25\\
  Snacks                    &16    &20.731  &1.235    &11.223 &1.236  &0&0&74&6 &25&25 \\
  Alcoholic beverages       &27    &110.706 &2.483    &45.196 &2.864  &0&0&350&30 &25&12\\
  Non-alcoholic beverages   &32    &10.627  &1.313    &6.113  &1.469  &0&0&48&9 &25&12\\
  \hline
\end{tabular}

  \end{scriptsize}
  \caption{Overview of data from \restA per category.}
  \label{tab:itemsRestA}
\end{table}

\begin{table}[!ht]
  \centering
  \begin{scriptsize}
    % !TeX spellcheck = en_US

\begin{tabular}{S S S S S S S SSSSS[table-format=2]  }
  \hline
    {\# time series} &{meanH} & {meanL} & {sdH} & {sdL} & {minH} & {minL} & {maxH} & {maxL} &{missH} & {missL}\\
  \hline
   30     &651.973  &19.587    &765.046  &38.178  &0&0&9769&184 &231&231\\
  \hline
\end{tabular}

  \end{scriptsize}
  \caption{Overview of data from \restB.}
  \label{tab:itemsRestB}
\end{table}

\begin{figure*}[!ht]
  \begin{minipage}[t]{0.45\linewidth}
    \centering
    \includegraphics[width=\textwidth]{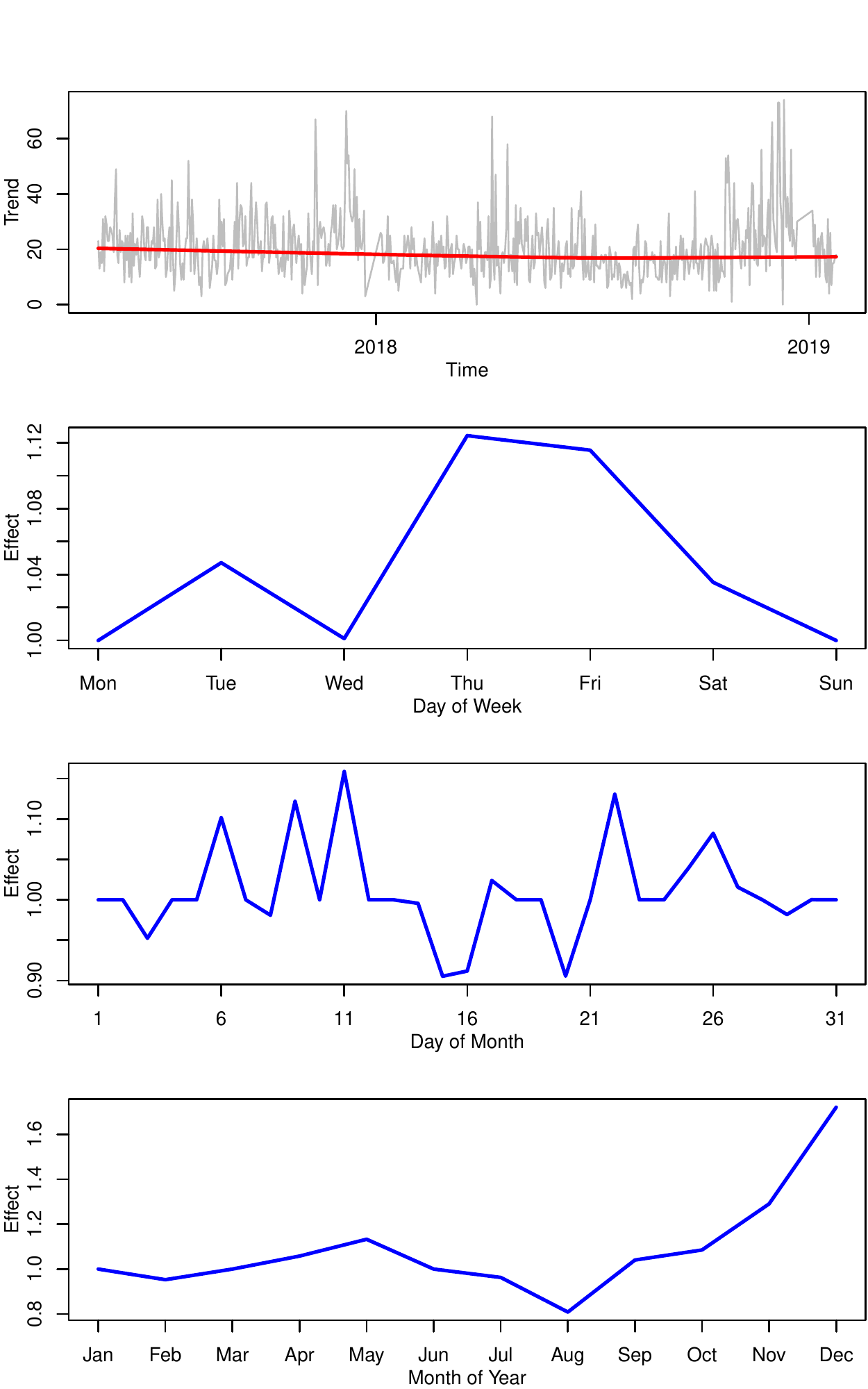}
    \caption{Structural characteristics of time series A1.}
    \label{A1char}
  \end{minipage}
  \hspace{0.9cm}
  \begin{minipage}[t]{0.45\linewidth}
    \centering
    \includegraphics[width=\textwidth]{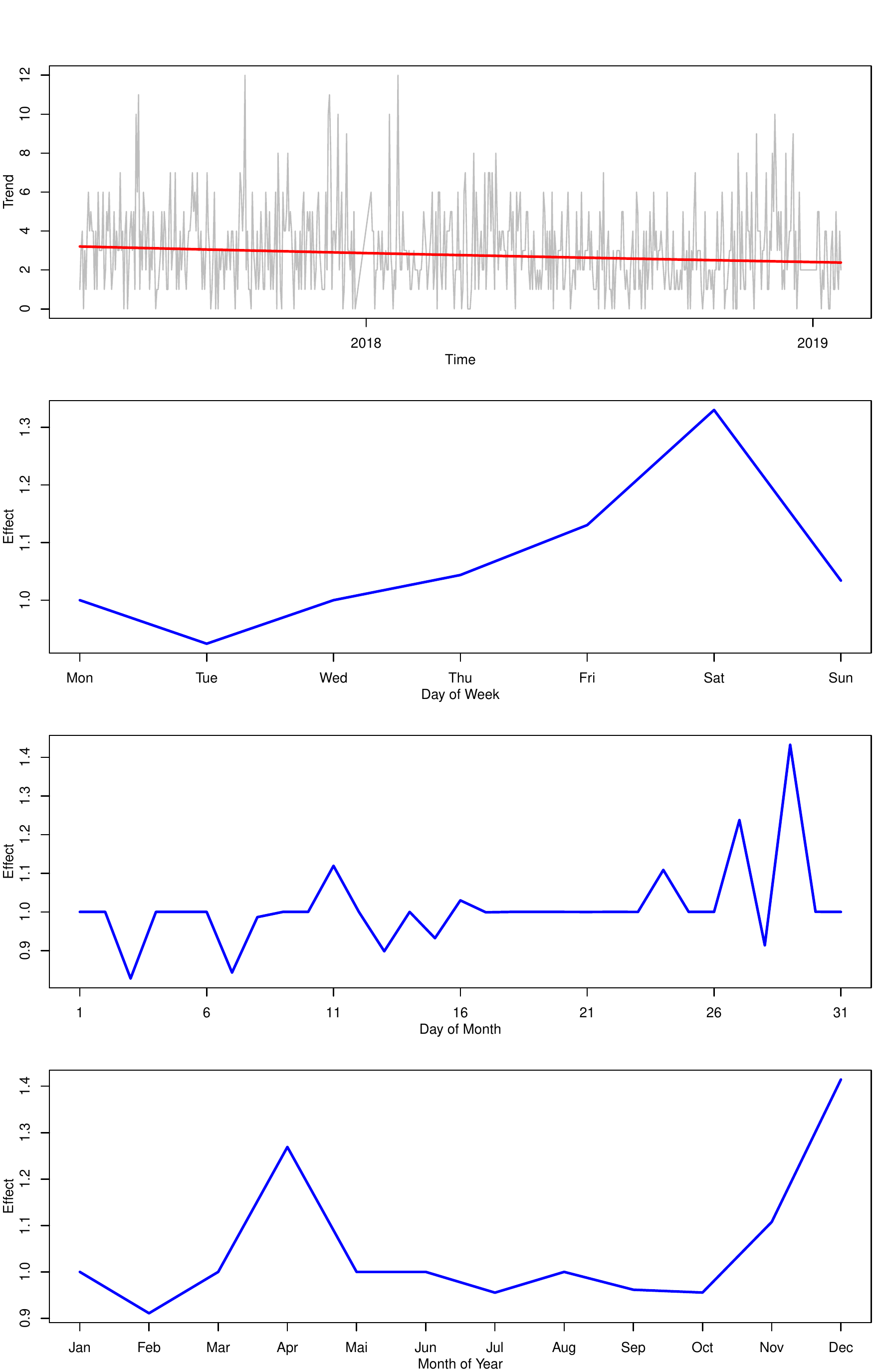}
    \caption{Structural characteristics of time series A2.}
    \label{A2char}
  \end{minipage}
  \begin{minipage}[t]{0.45\linewidth}
    \centering
    \includegraphics[width=\textwidth]{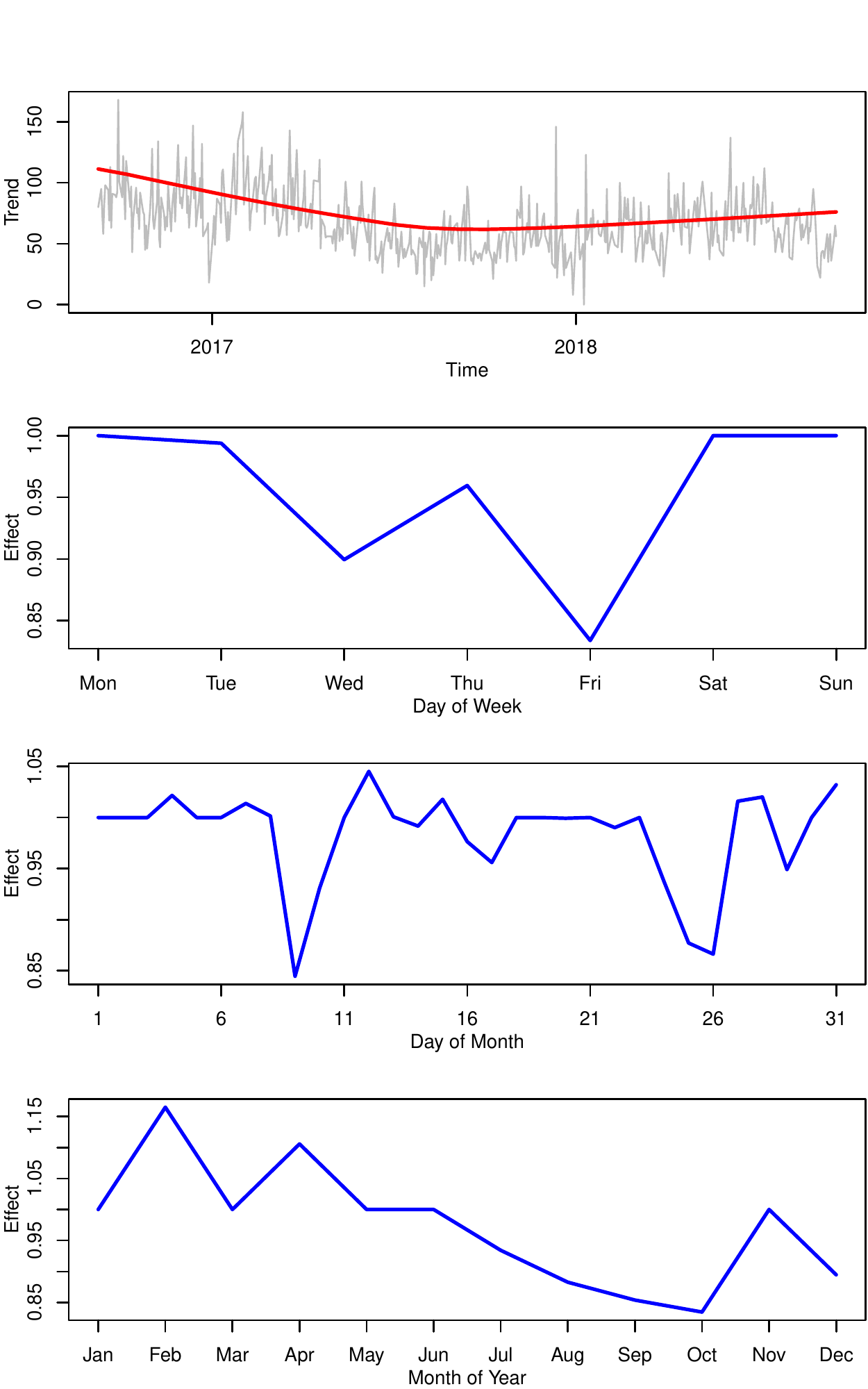}
    \caption{Structural characteristics of time series B1.}
    \label{B1char}
  \end{minipage}
  \hspace{0.9cm}
  \begin{minipage}[t]{0.45\linewidth}
    \centering
    \includegraphics[width=\textwidth]{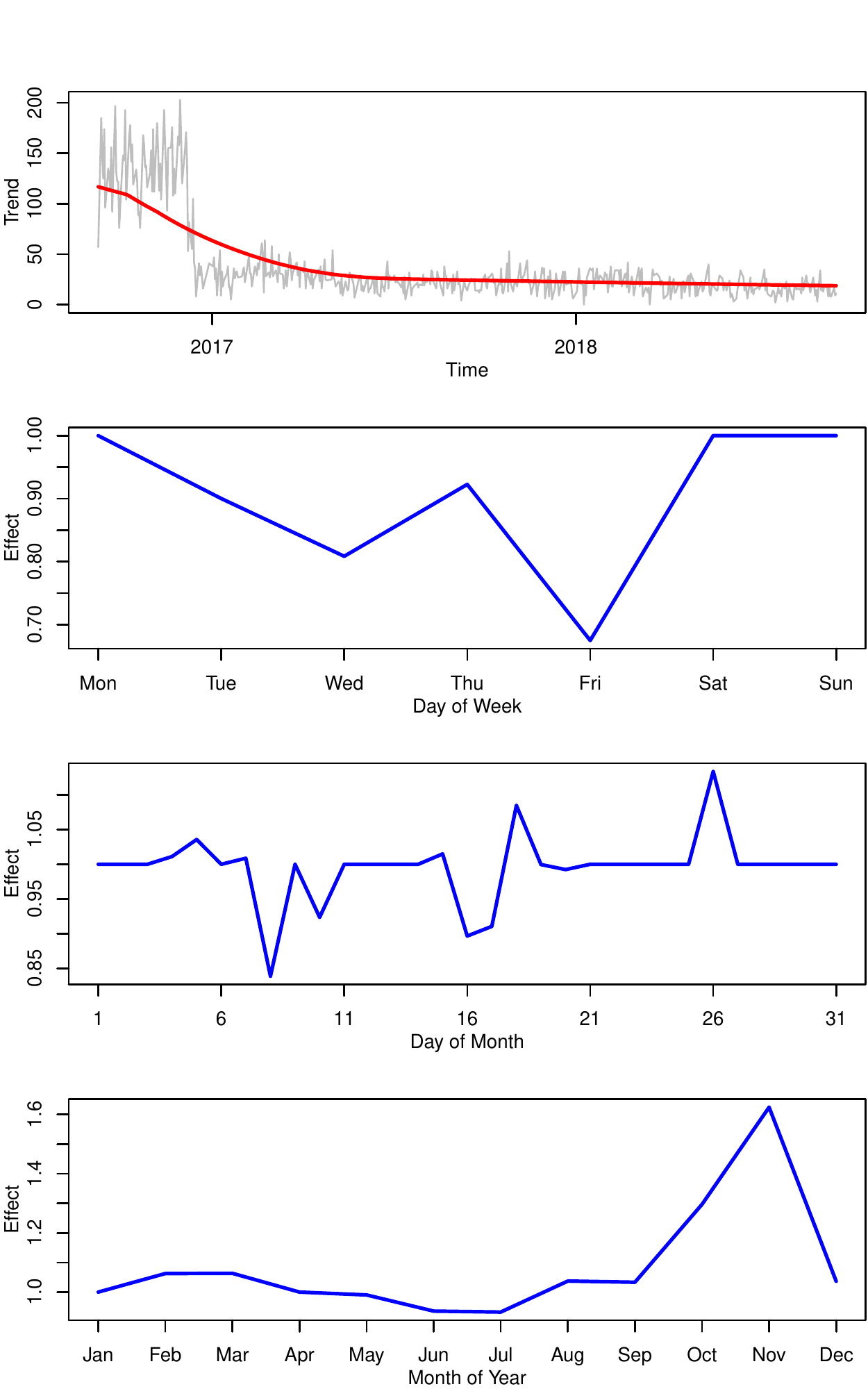}
    \caption{Structural characteristics of time series B2.}
    \label{B2char}
  \end{minipage}
\end{figure*}

In order to provide some structural characteristics of the POS data four representative time
series A1, A2, B1, and  B2 are considered. A1 and A2 correspond to \restA,
while B1, and B2 belong to \restB.
In   Figures \ref{A1char}   --  \ref{A2char}, the time series are decomposed into
trend and weekly, monthly, and yearly seasonal patterns.
The decomposition is obtained from the proposed model that assumes a negative binomial
target, compare to Section \ref{sec:negBinomMod}.
This model is chosen for the decomposition since it achieved the best results in the
performance evaluation presented in Section \ref{sec:eval}.
Please recall that in this model the seasonal effects (for weekday, day of month, month of year)
act multiplicatively on the trend. Note, that the seasonal effects are obtained
by passing the coefficients of the seasonality function to the exponential function.
Investigation of the Figures \ref{A1char}  --  \ref{A2char} shows that the trend is often
quite constant (A1, A2), while at the same time also smooth (B1) and drastic (B2) trend
changes appear.
Moreover, weekly, monthly, and yearly seasonal patterns are observed.
For instance, time series A1 shows an increasing number of sales at the end of the year,
time series A2 shows a similar behavior at the end of the month, and the
time series B1 and B2 show the lowest number of sales on Fridays.
Finally, the presented time series also show some outliers.
It should be mentioned, that the outliers observed in these time series
are not as strong as outliers we could observe in other time series
from \restA, or \restB.
For reasons of brevity we
do not include further time series plots in this section.

\subsection{Implementation of the Models}
\label{ssec:implementation}

We have implemented the models proposed in Section \ref{sec:methodology}
using \RStan \cite{rstan2018},
the interface  between the programming languages \R \cite{RVienna}
and \Stan \cite{Stan2017}.
\Stan is known as a state-of-the-art platform for statistical modeling and
high-performance statistical computation.
In particular, \Stan can be used to compute the MAP-estimate of the model parameters
and also to perform full Bayesian inference via \textit{Markov chain Monte Carlo}  (MCMC) sampling.
Hence, a few lines of  \Stan code,
complemented by a \R script to compute the matrices $\mathbf{X}$ and $\mathbf{Z}$,
suffice to express our Bayesian models.

For the remainder of this section, we use the following abbreviations to identify our models:
In case that the target values are assumed to be negative binomial distributed our
method is abbreviated by \negBinomModel and in case that a
normal distribution is assumed our  method is denoted as \normalModel.

\subsection{Insights on the Proposed Models}
\label{ssec:eval1}
In this subsection, we illustrate how well the proposed models fit the properties of
the provided real-world data.
The evaluation is based on representative datasets from the above described
restaurants A and B. In Section \ref{standardAppr}, the priors given by the
Equations   \eqref{eq:priorNormal} -- \eqref{eq:normalPriorCoeff} are assigned to the trend functions of the considered models.
Further, the  standard settings  specified in \eqref{eq:standardSettings} are assigned
to the tuning parameters of the models. These are our recommendations in case
that the model optimization has to take place automatically and at low
computational cost. In Section \ref{advancedAppr}, we illustrate the superior
performance of more advanced approaches for a time series associated with \restB.

\subsubsection{Standard Approaches}
\label{standardAppr}
\begin{figure*}[!ht]
  \centering
  \begin{minipage}[t]{0.45\linewidth}
    \centering
    \includegraphics[width=\textwidth]{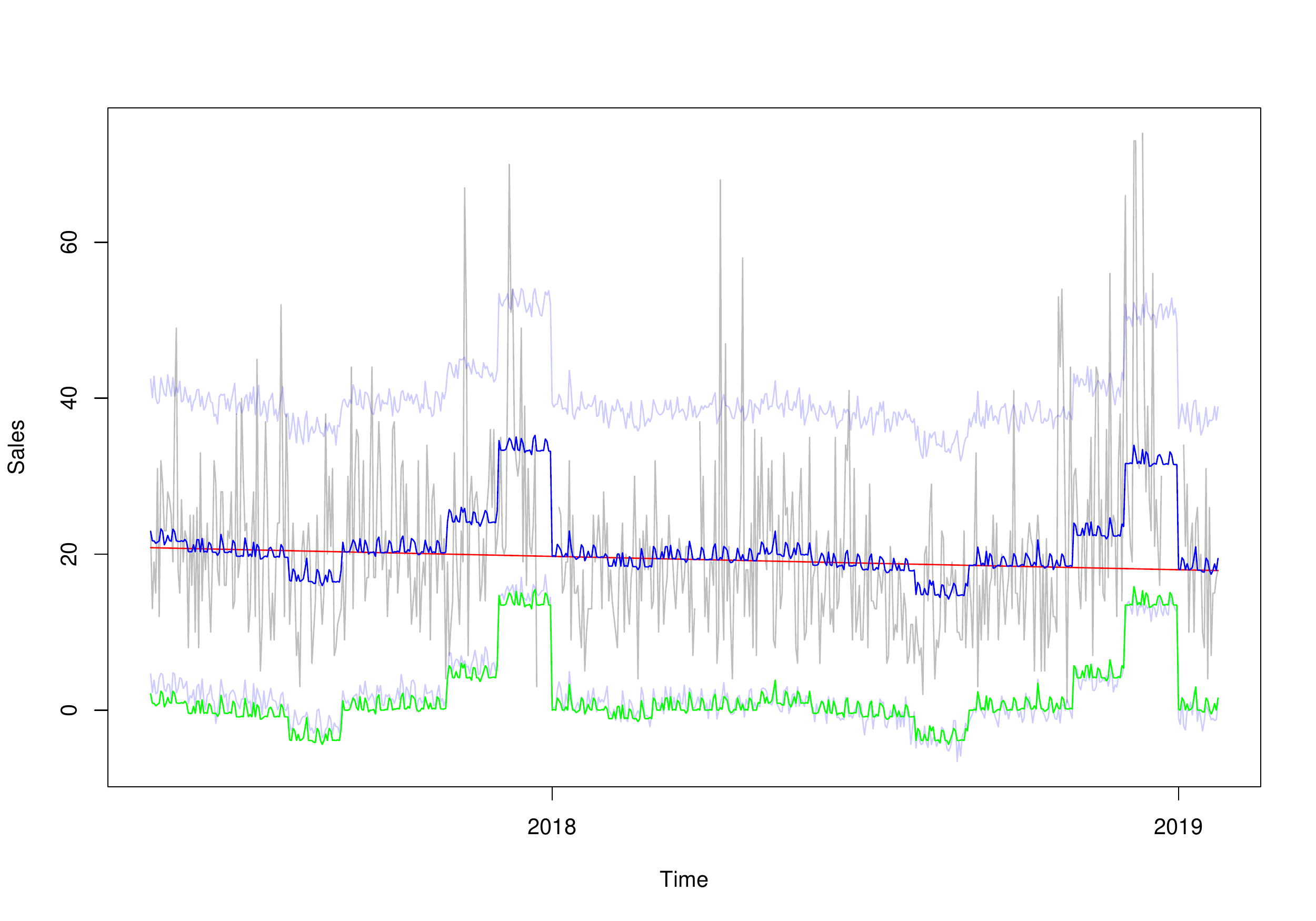}
    \caption{Model fit of \normalModel on the time series A1.}
    \label{T1Gauss}
  \end{minipage}
  \hspace{0.9cm}
  \begin{minipage}[t]{0.45\linewidth}
    \centering
    \includegraphics[width=\textwidth]{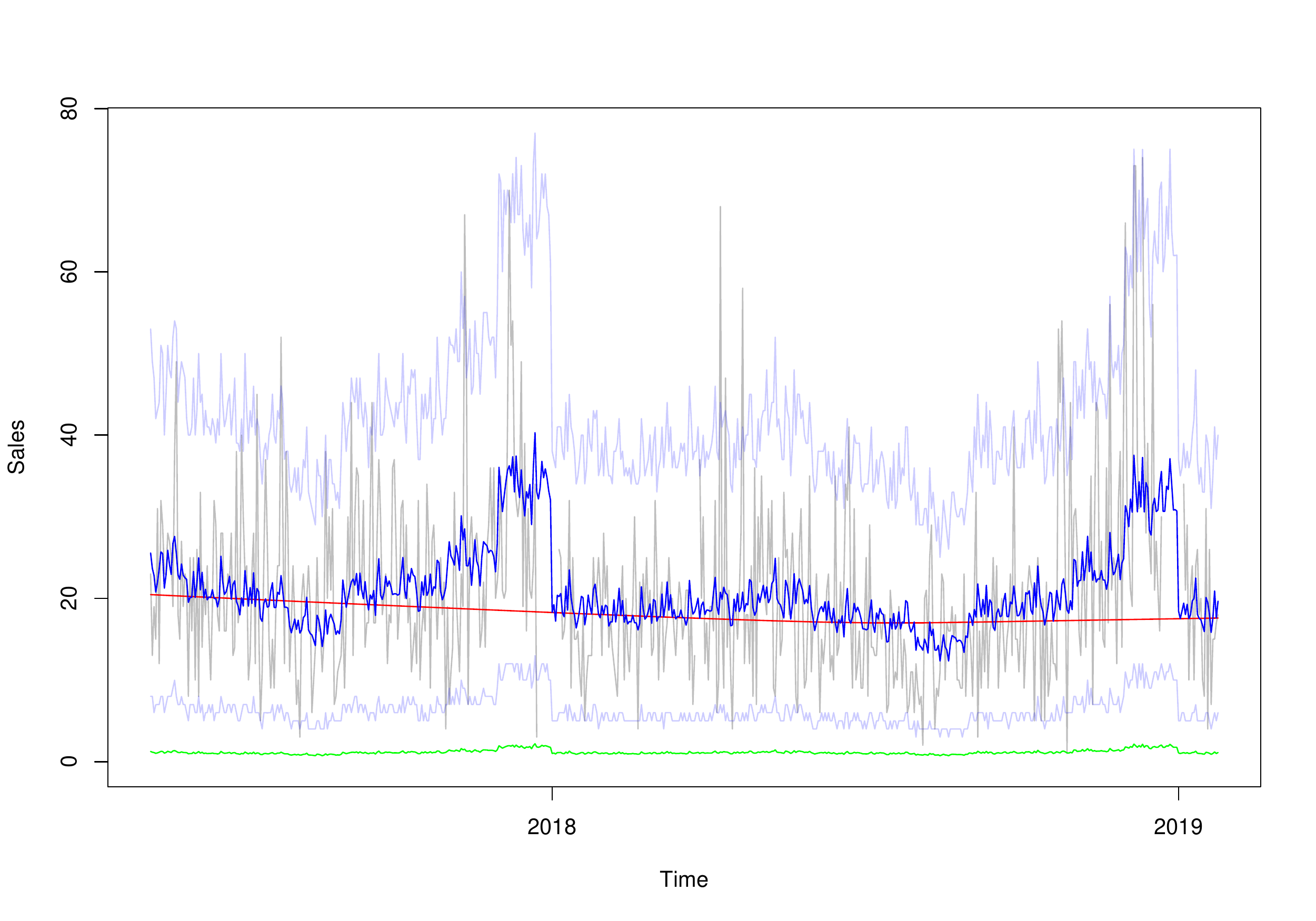}
    \caption{Model fit of \negBinomModel on the time series A1.}
    \label{T1NegBinom}
  \end{minipage}
\end{figure*}

\begin{figure*}[!ht]
  \centering
  \includegraphics[width=\textwidth]{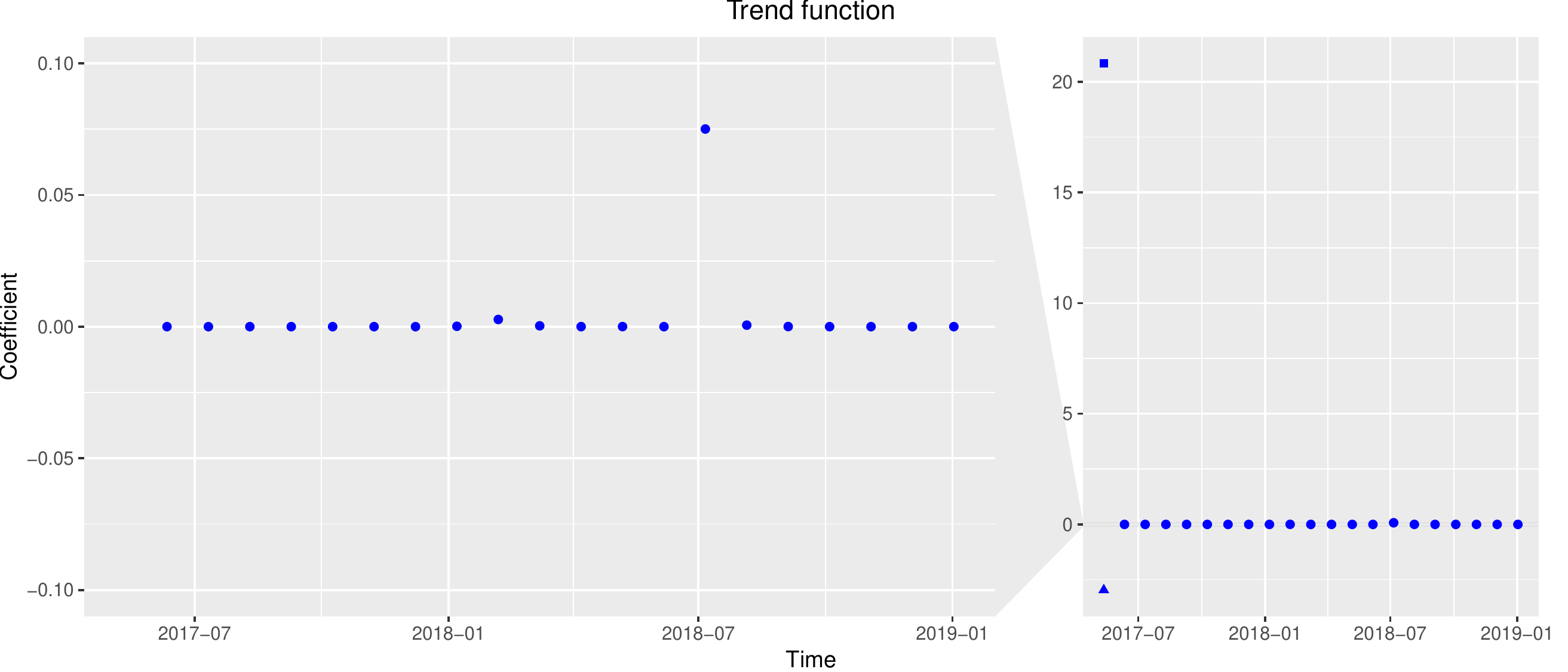}
  \caption{Coefficients of the trend function (\normalModel, time series A1).}
  \label{A1TrendGauss}
  \vspace{0.5cm}
  \includegraphics[width=\textwidth]{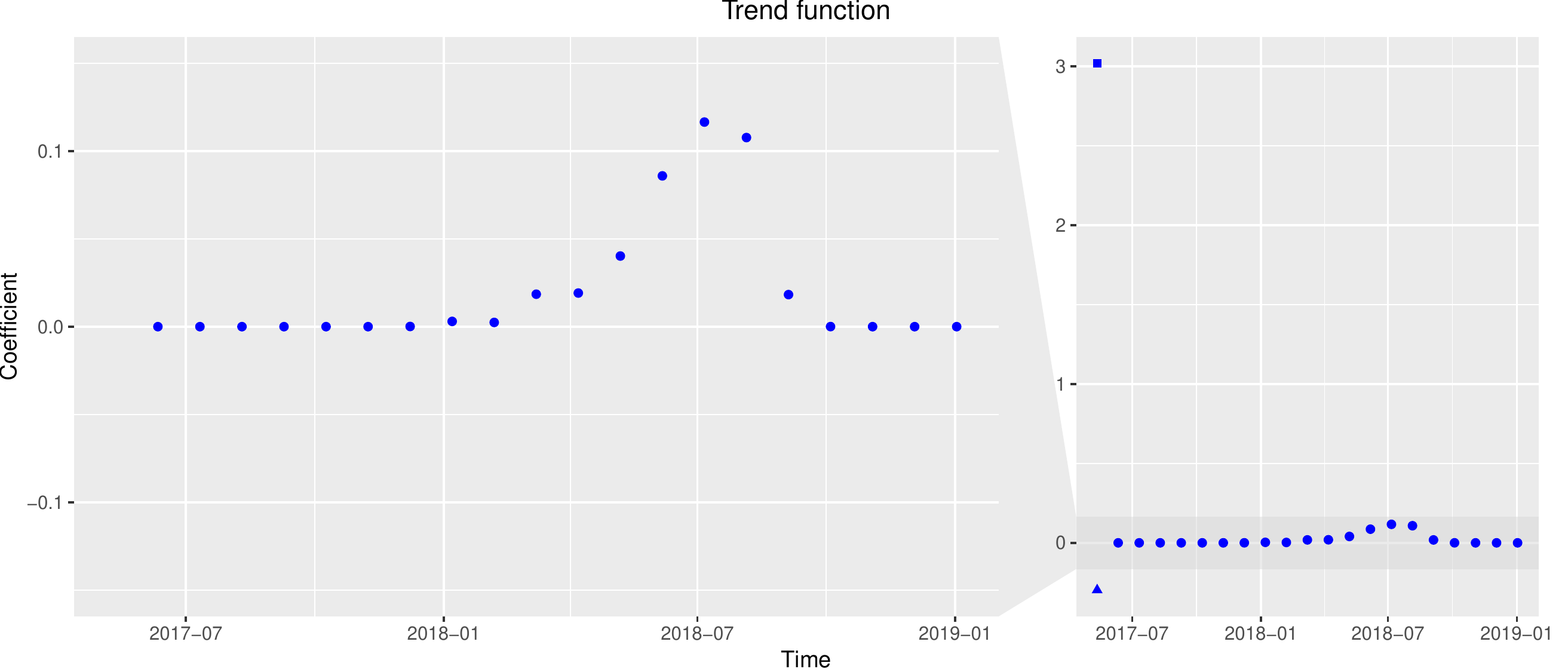}
  \caption{Coefficients of the trend function (\negBinomModel, time series A1).}
  \label{A1TrendNegBinom}
\end{figure*}

\begin{figure*}[!ht]
  \includegraphics[width=\textwidth]{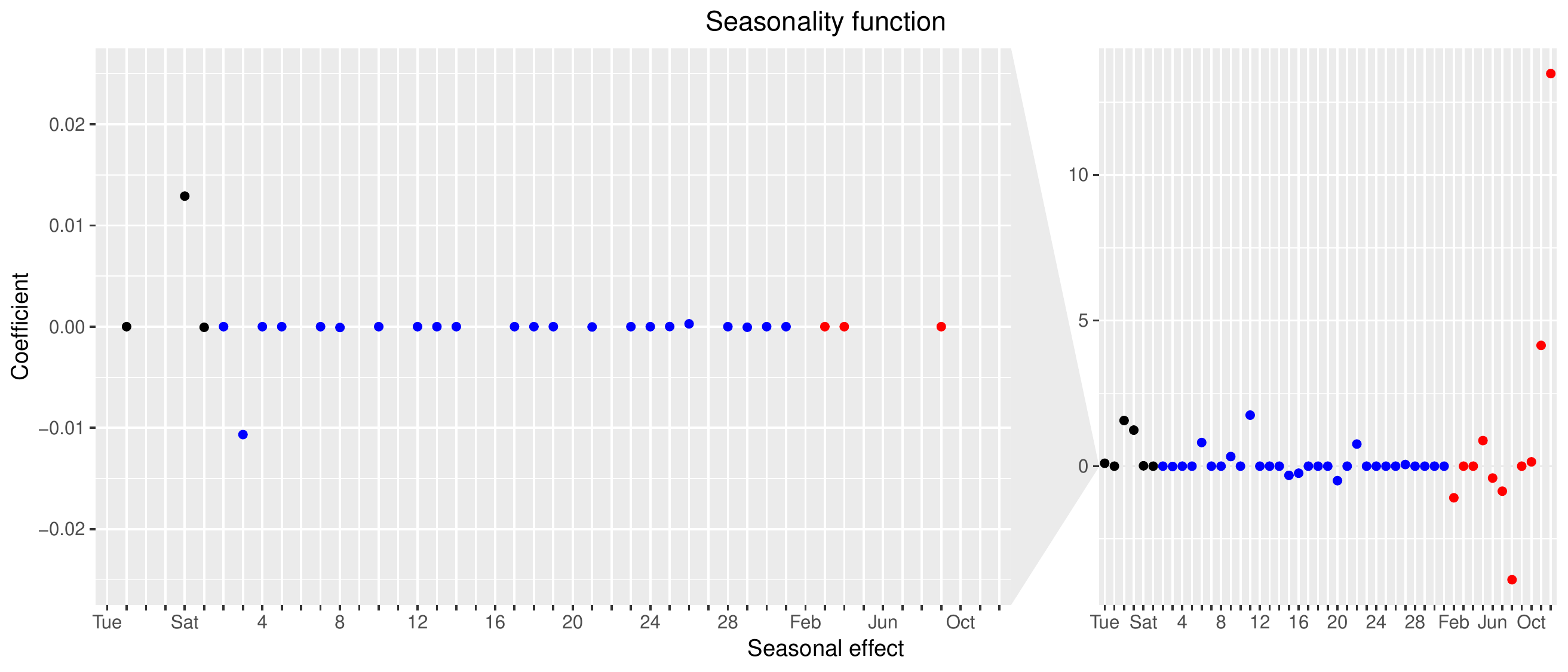}
  \caption{Coefficients of the seasonality function (\normalModel, time series A1).}
  \label{A1SeasonGauss}
  \vspace{0.5cm}
  \includegraphics[width=\textwidth]{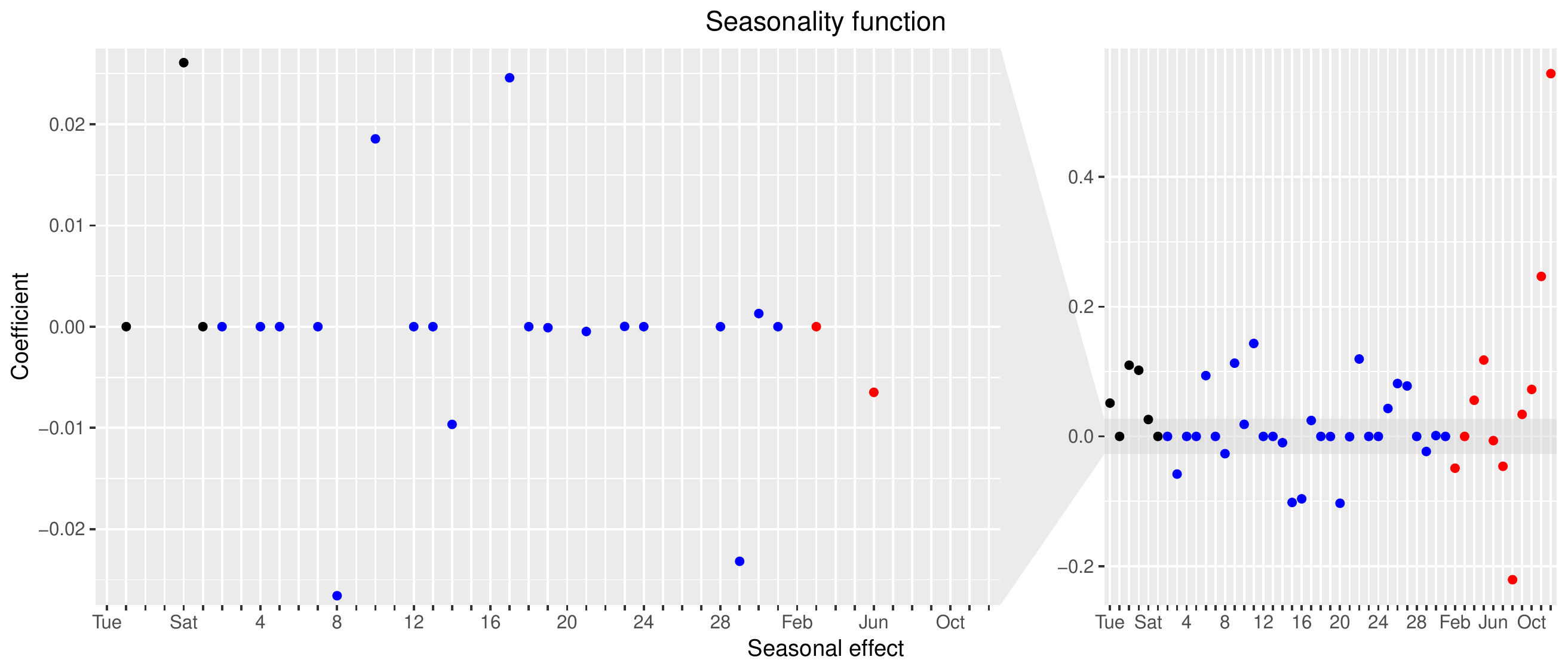}
  \caption{Coefficients of the seasonality function (\negBinomModel, time series A1).}
  \label{A1SeasonNegBinom}
\end{figure*}

At first, a representative time series A1 (compare to Section \ref{ssec:data}) corresponding to \restA is
considered. In Figure \ref{T1Gauss}, the model fit is visualized for \normalModel
\eqref{eq:normalModel} and in Figure \ref{T1NegBinom}, the model fit of \negBinomModel is shown.
The trend function is plotted in red, the seasonal function is colored green,
the expected value $\mathbb{E}(\mathbf{y}|\widehat{\boldsymbol\beta},\widehat{\boldsymbol\gamma})$, see  Equation  \eqref{eq:expectedValue2},
is drawn in dark blue and, finally, the $95\%$ prediction intervals for the
components of $\mathbf{y}$ are plotted in light blue. In particular, for \negBinomModel,
the exponential function of the trend and the season are plotted.
Compared to  \normalModel, the seasonal function of \negBinomModel
takes small values. The reason for this is that the seasonal function acts
in a multiplicative way on the trend, $\mu_t = \operatorname{exp}(g(\delta(t)))\operatorname{exp}(s(t))$,
in \negBinomModel and in an additive way, $\mu_t = g(\delta(t))+s(t)$, in \normalModel.
For \normalModel, $94.63\%$ of the observed sales lie within the $95\%$ prediction intervals.
The prediction intervals of \negBinomModel cover $94.97\%$ of the sales. Thus, both models provide a
reasonable fit of the observed time series. In Figures \ref{A1TrendGauss}
and \ref{A1TrendNegBinom} the coefficients of the trend functions are visualized.
In particular, each coefficient is plotted against the time point it has a
non-zero effect on the model for the first time. Since the coefficients of
the global polynomial effect the model from the beginning, different symbols are used
to visualize them (a square for the intercept $\gamma_1$, a triangle for the
coefficient $\gamma_2$).   While \normalModel does not consider any significant
trend changes, \negBinomModel detects a slight change in the second
half of the year 2019. The coefficients of the seasonality functions are shown
in  Figures \ref{A1SeasonGauss} and \ref{A1SeasonNegBinom}. While \negBinomModel
considers more seasonal effects as significant than \normalModel,
both models agree on the effects with the highest impact
(December, November, August, Thursday, Friday, eleven-th day of month, \ldots).

\begin{figure*}[!ht]
  \centering
  \begin{minipage}[t]{0.45\linewidth}
    \centering
    \includegraphics[width=\textwidth]{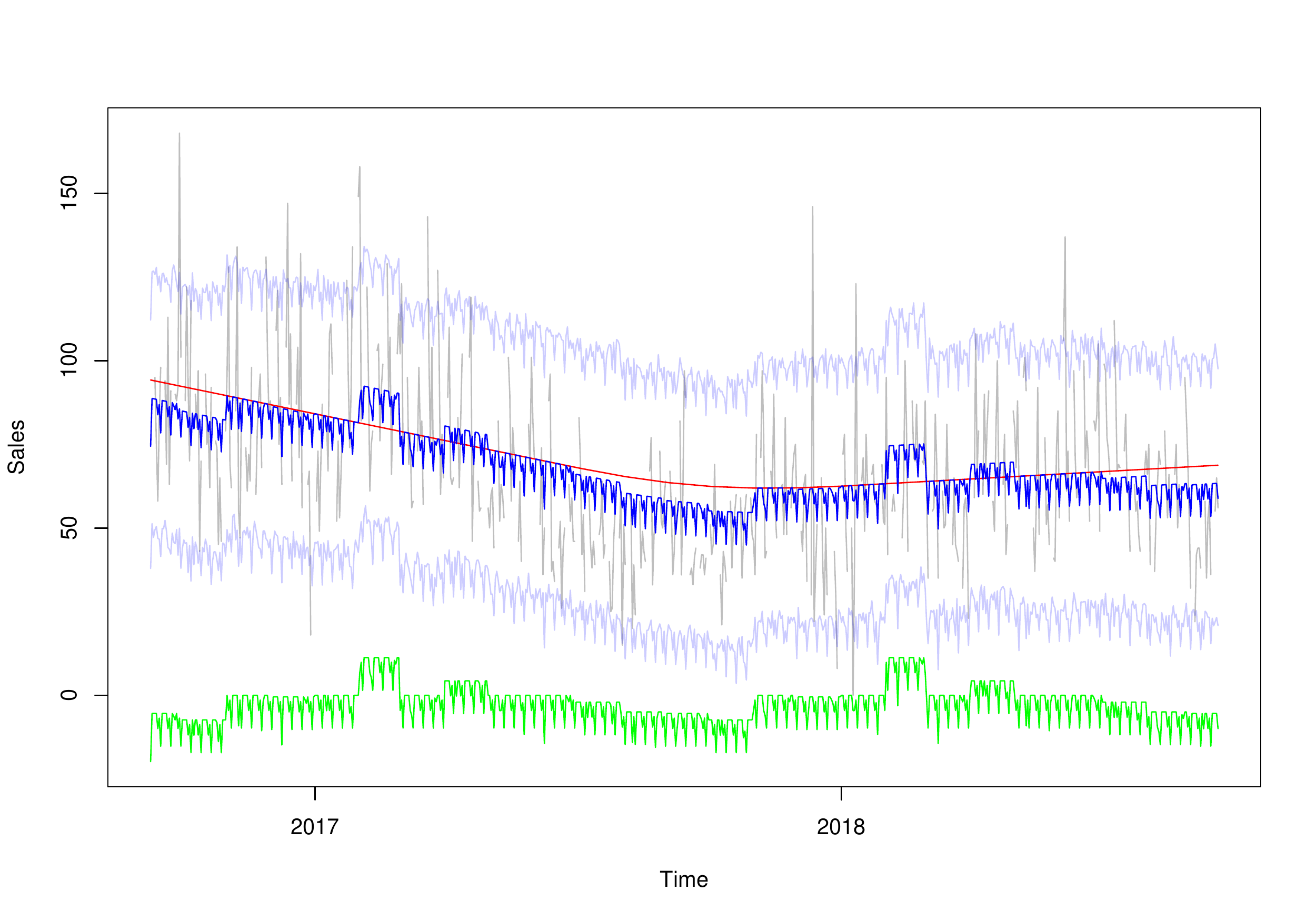}
    \caption{Model fit of \normalModel on the time series B1.}
    \label{B1Gauss}
  \end{minipage}
  \hspace{0.9cm}
  \begin{minipage}[t]{0.45\linewidth}
    \centering
    \includegraphics[width=\textwidth]{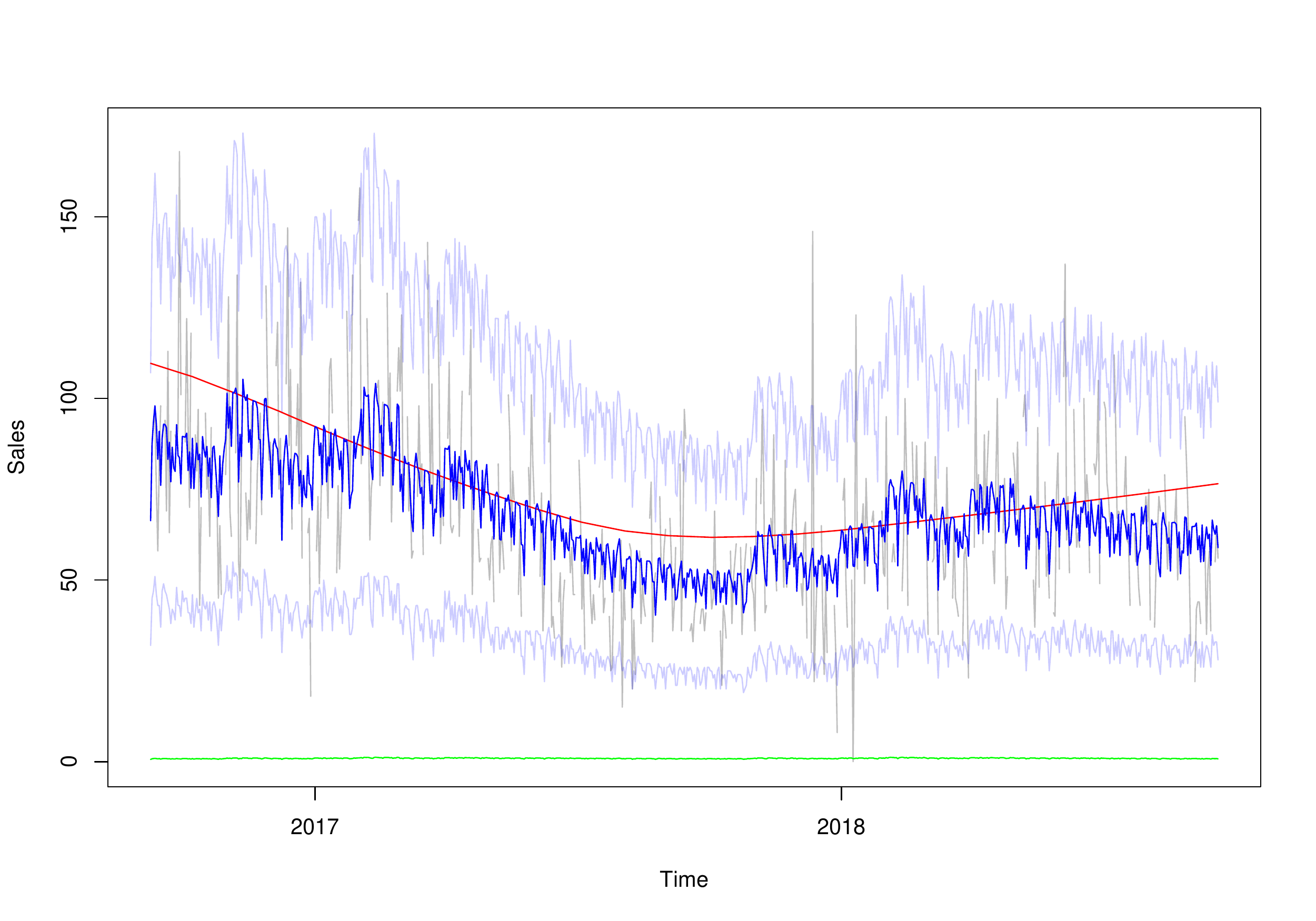}
    \caption{Model fit of \negBinomModel on the time series B1.}
    \label{B1NegBinom}
  \end{minipage}
\end{figure*}

\begin{figure*}[!ht]
  \centering
  \includegraphics[width=\textwidth]{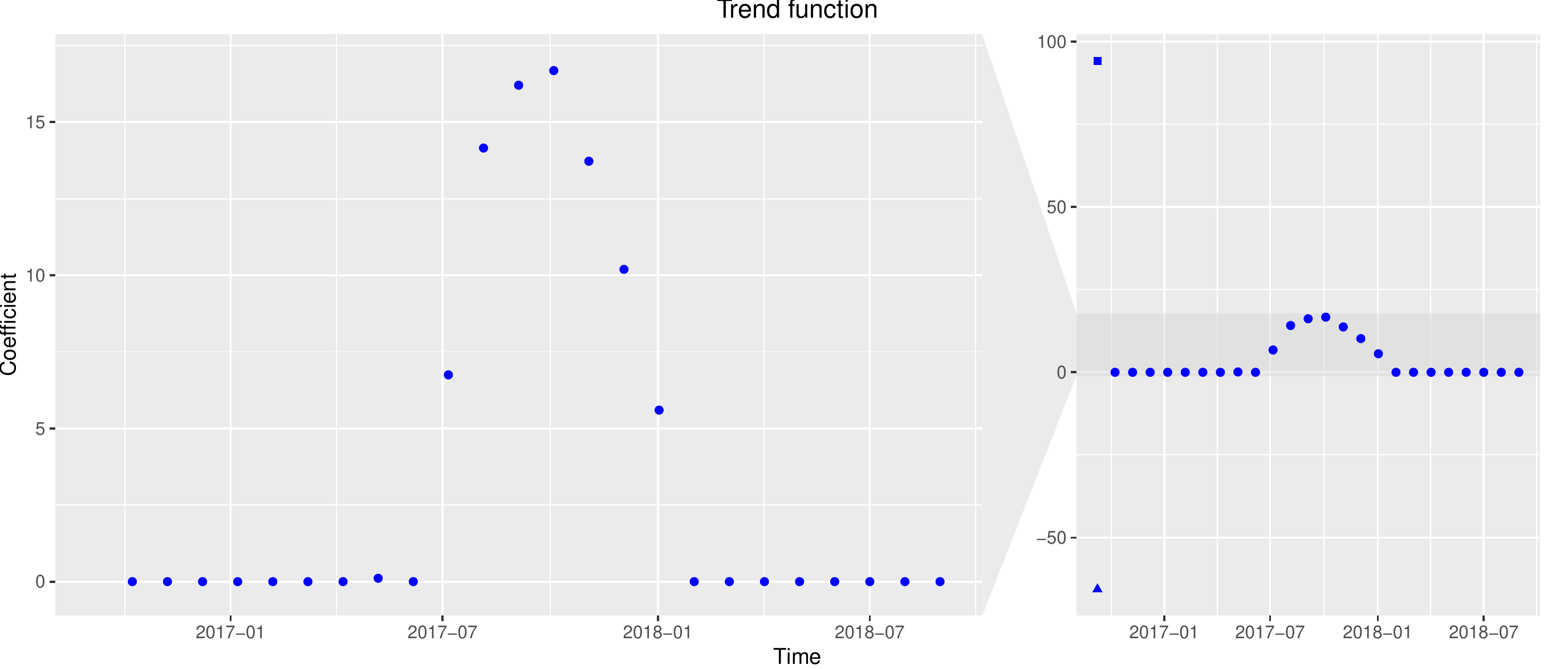}
  \caption{Coefficients of the trend function (\normalModel, time series B1).}
  \label{B1TrendGauss}
  \vspace{0.5cm}
  \includegraphics[width=\textwidth]{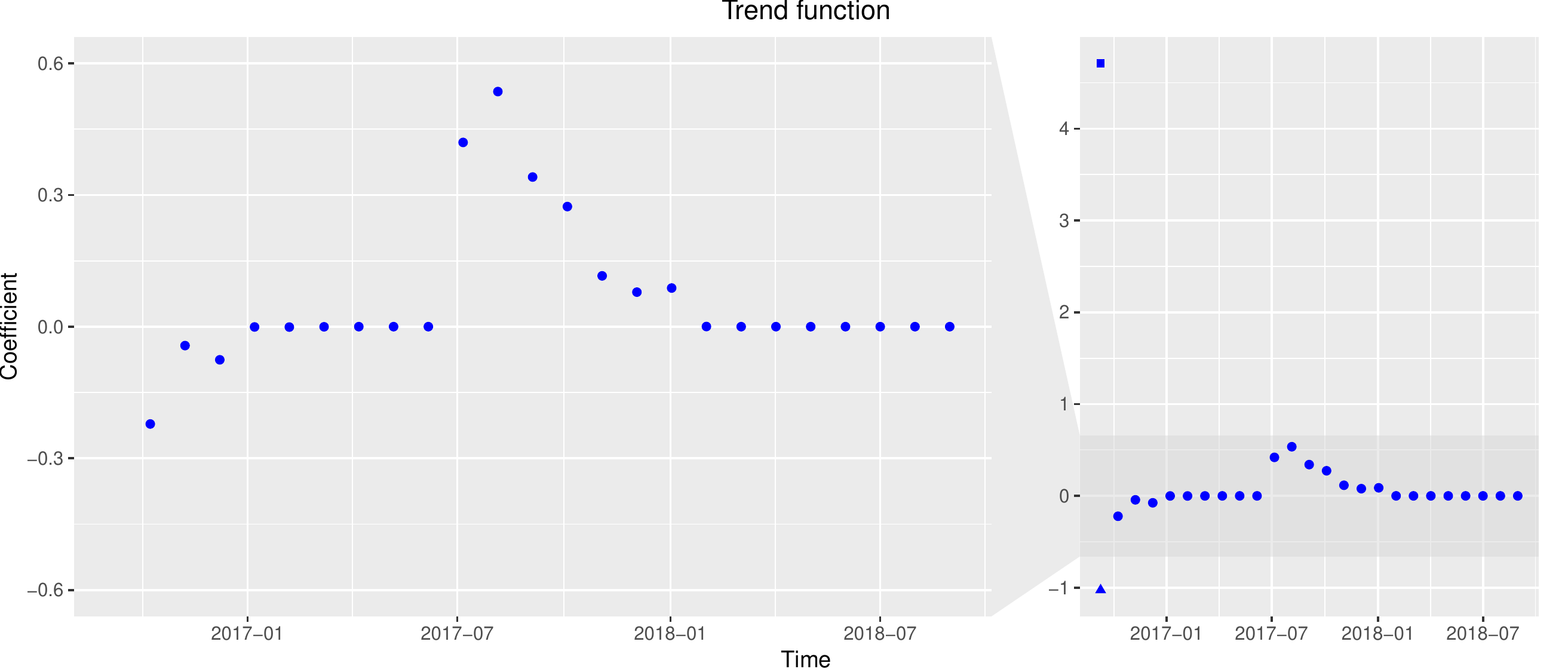}
  \caption{Coefficients of the trend function (\negBinomModel, time series B1).}
  \label{B1TrendNegBinom}
\end{figure*}

\begin{figure*}[!ht]
  \centering
  \includegraphics[width=\textwidth]{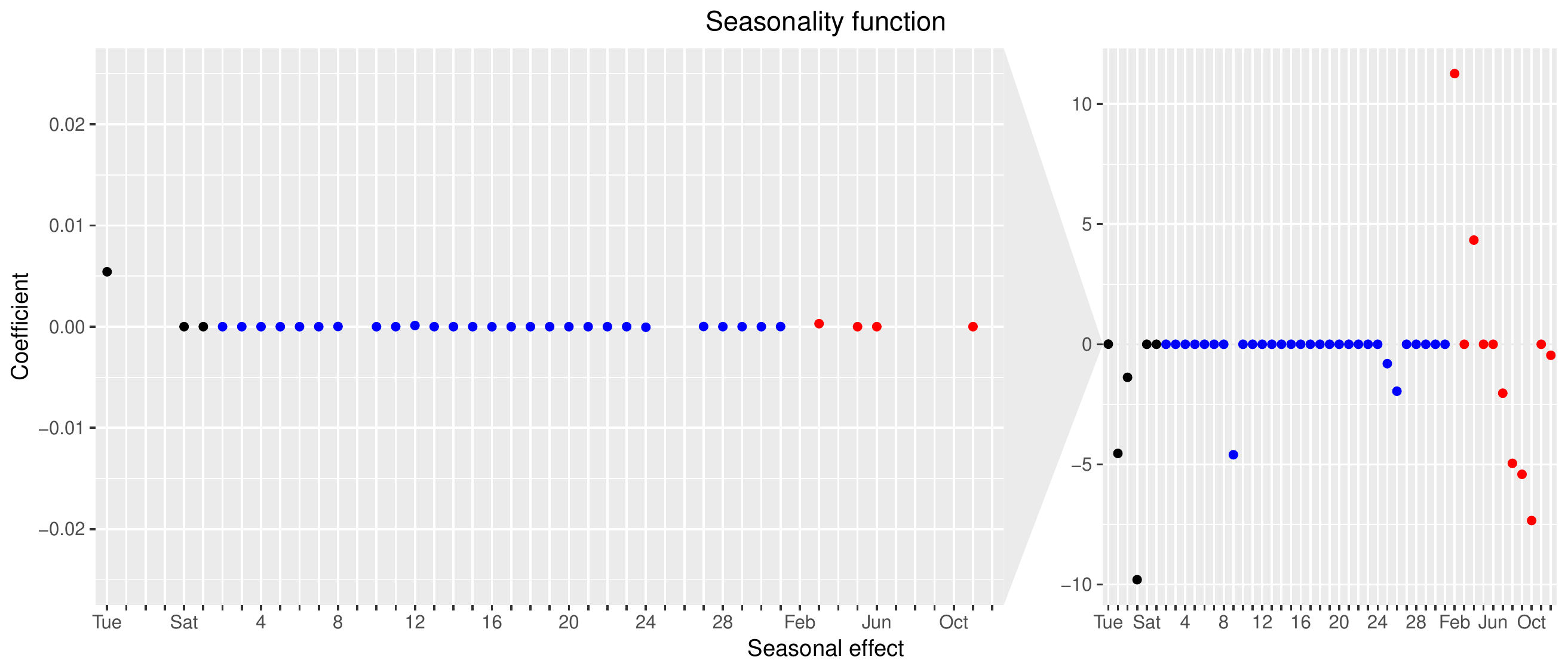}
  \caption{Coefficients of the seasonality function (\normalModel, time series B1).}
  \label{B1SeasonGauss}
  \vspace{0.5cm}
  \includegraphics[width=\textwidth]{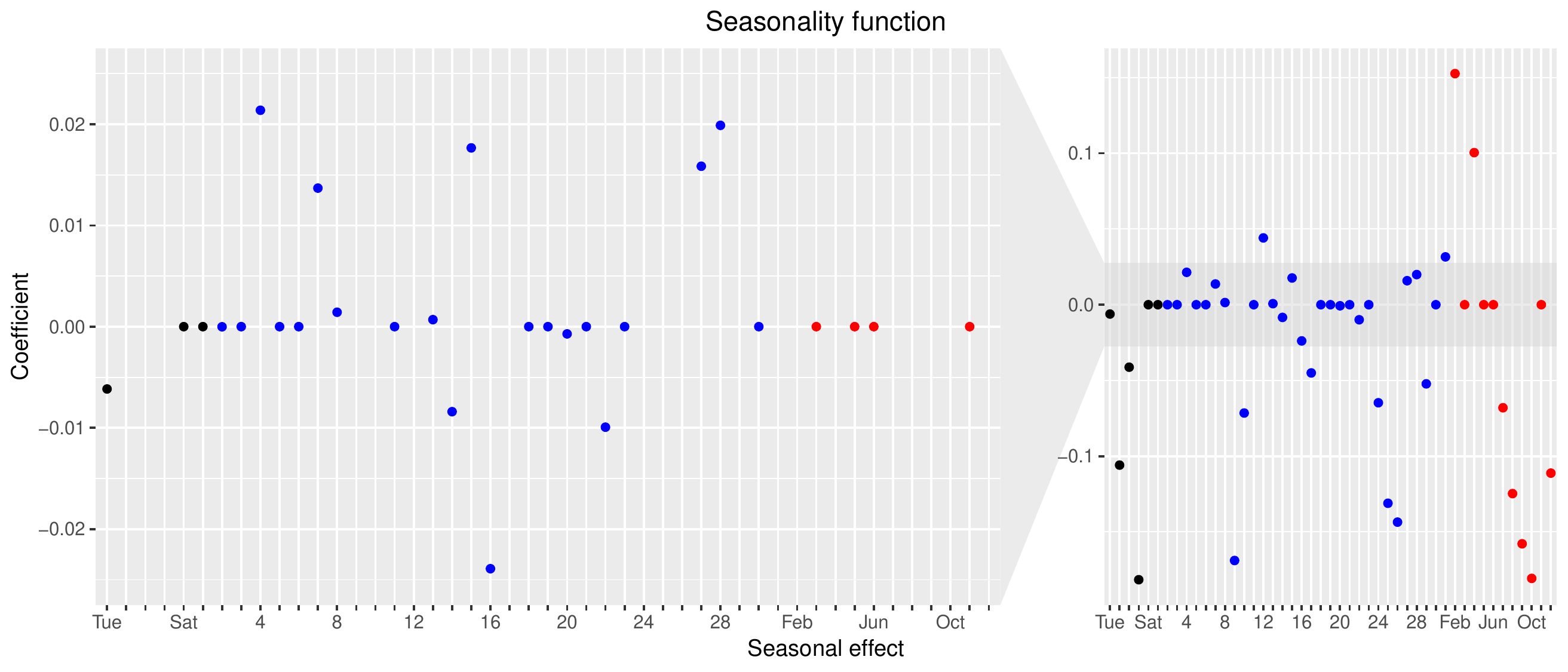}
  \caption{Coefficients of the seasonality function (\negBinomModel, time series B1).}
  \label{B1SeasonNegBinom}
\end{figure*}

Now a representative time series B1 (compare to Section \ref{ssec:data}) belonging to \restB is considered.
In  Figures \ref{B1Gauss} -- \ref{B1SeasonNegBinom} the model fit is shown
for \normalModel and for \negBinomModel
The figures can be interpreted in the same way it has been done for time series A1.
For \normalModel, $93.92157\%$ of the observed sales lie within the
corresponding $95\%$ prediction intervals. The prediction intervals of \negBinomModel
cover $95.29412\%$ of the sales.

\subsubsection{Advanced Approaches}
\label{advancedAppr}

\begin{figure*}[!ht]
  \centering

  \includegraphics[width=0.75\textwidth]{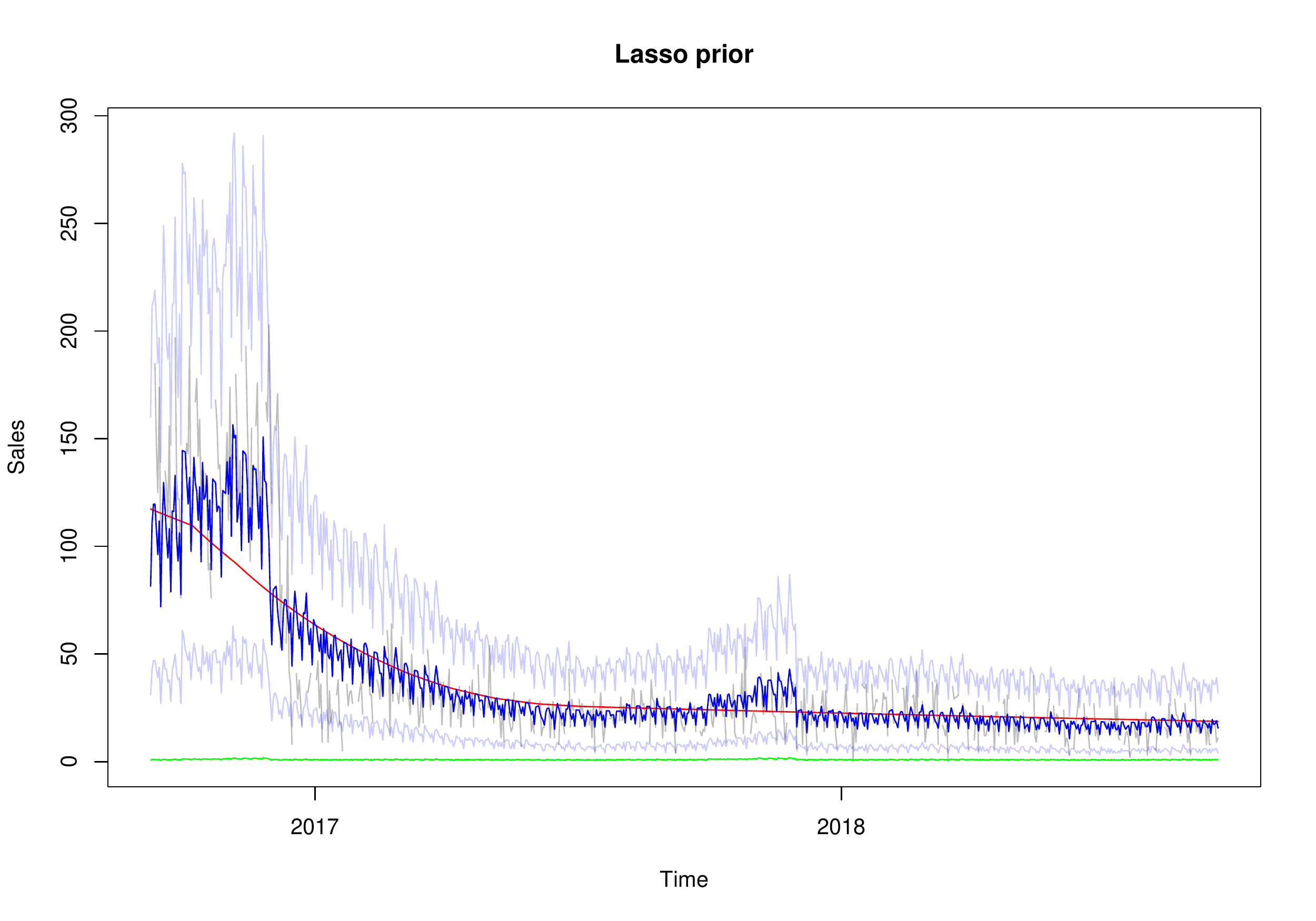}

  \includegraphics[width=0.75\textwidth]{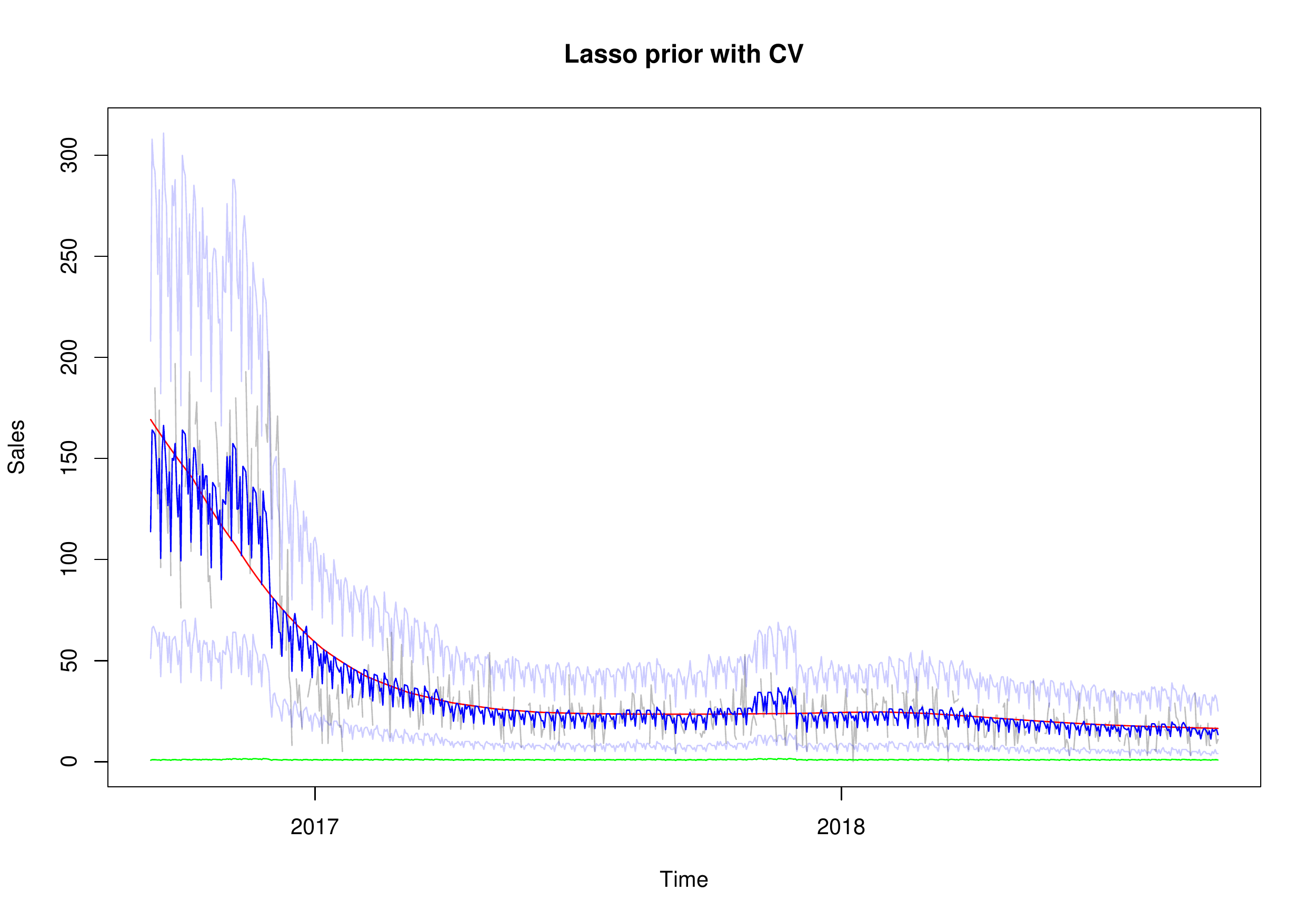}

  \includegraphics[width=0.75\textwidth]{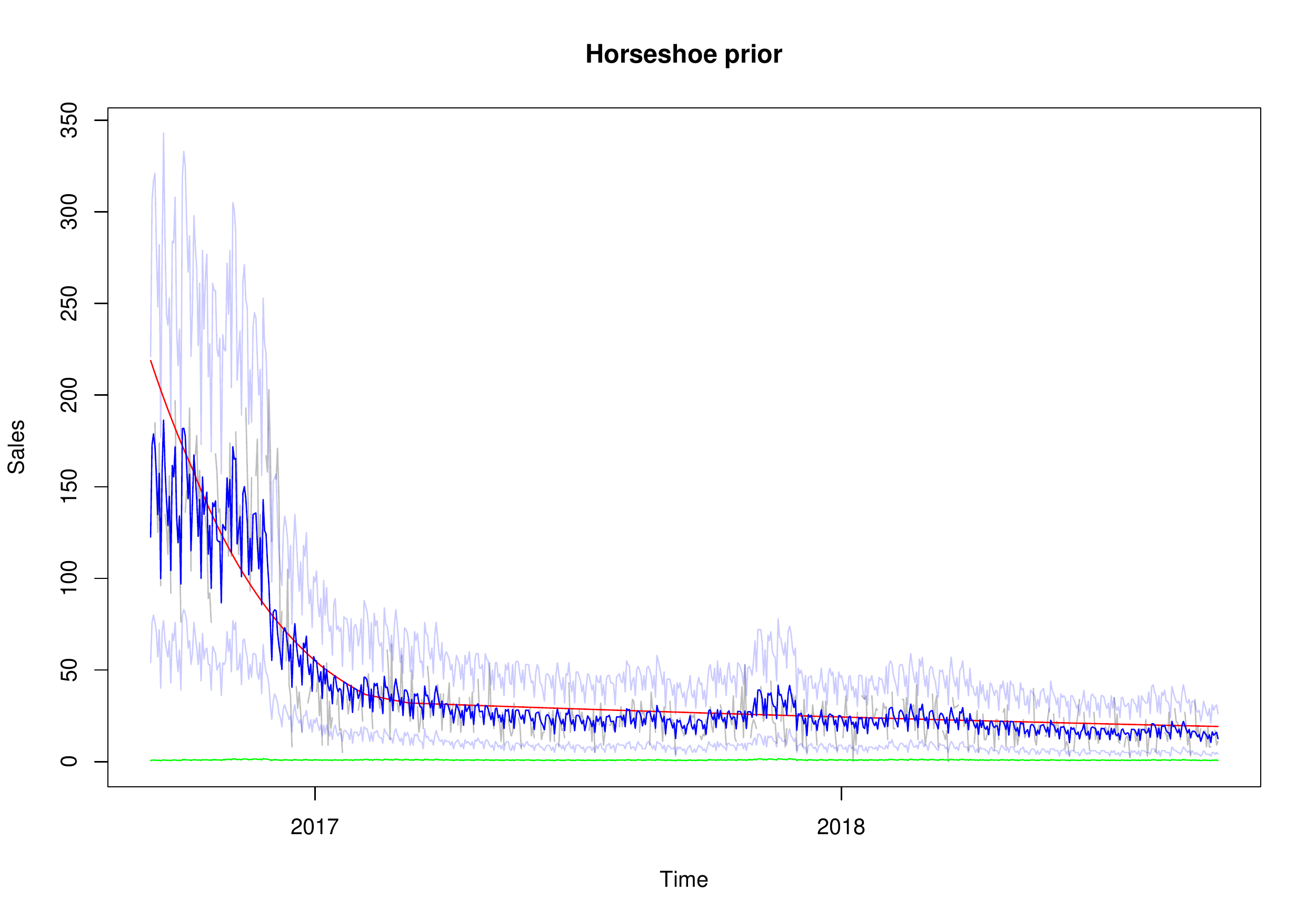}
  \caption{Model fit of \negBinomModel on the time series B2.}
  \label{B2ModelFitNegBinom}
\end{figure*}

\begin{figure*}[!ht]
  \centering
  \includegraphics[width=\textwidth]{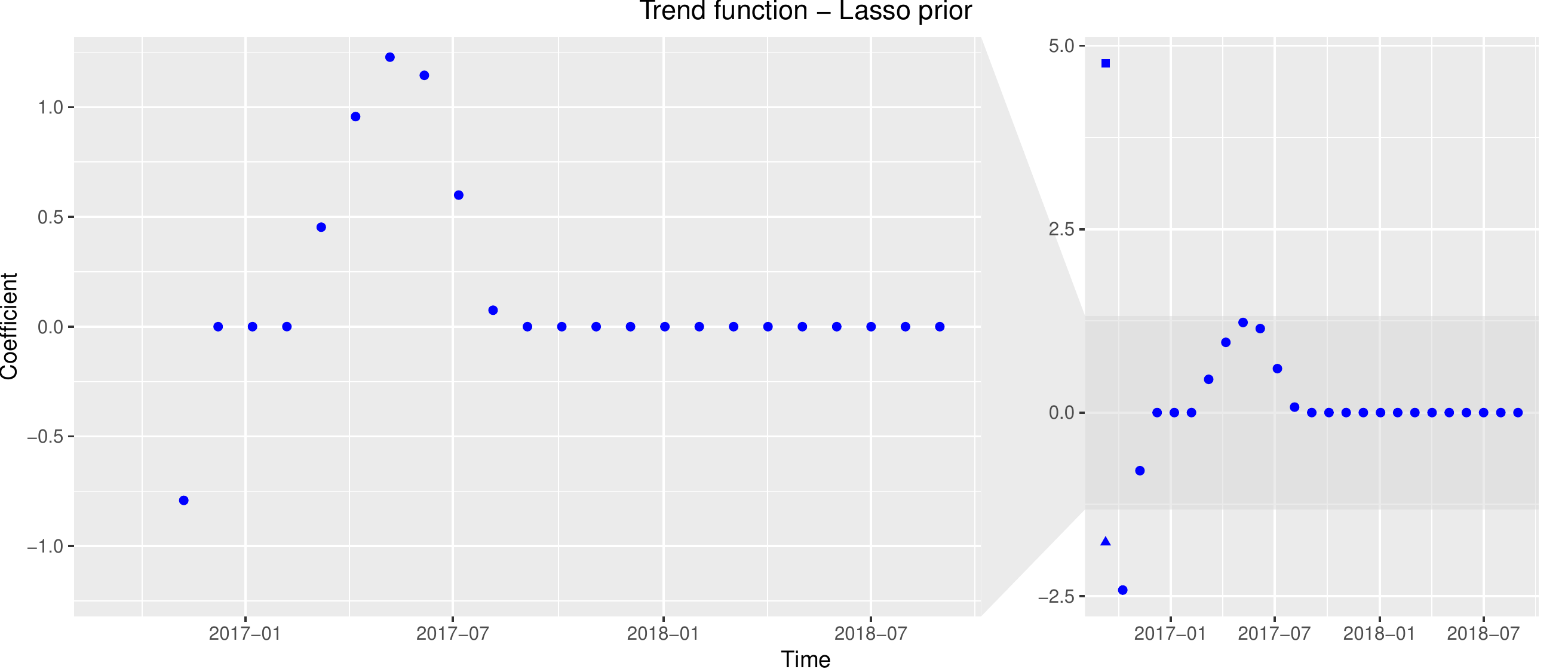}

  \vspace{0.5cm}
  \includegraphics[width=\textwidth]{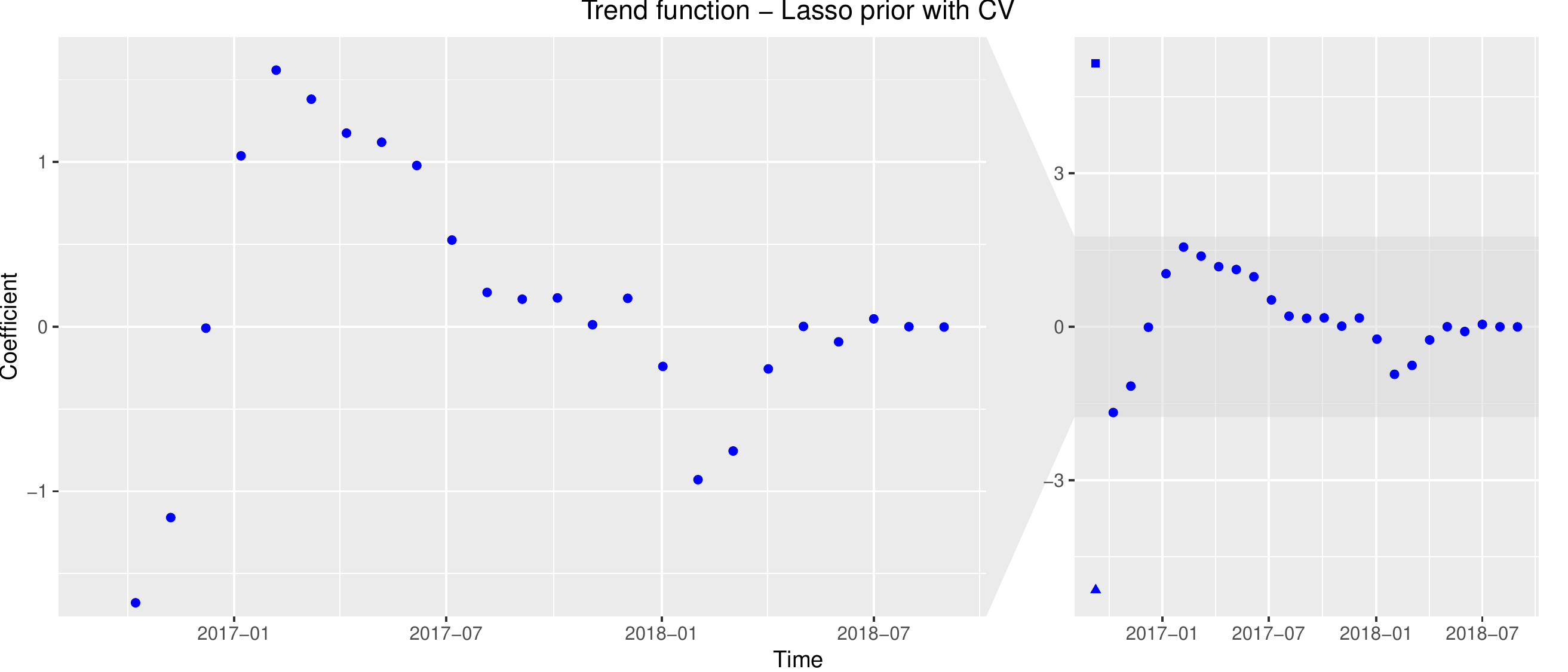}

  \vspace{0.5cm}
  \includegraphics[width=\textwidth]{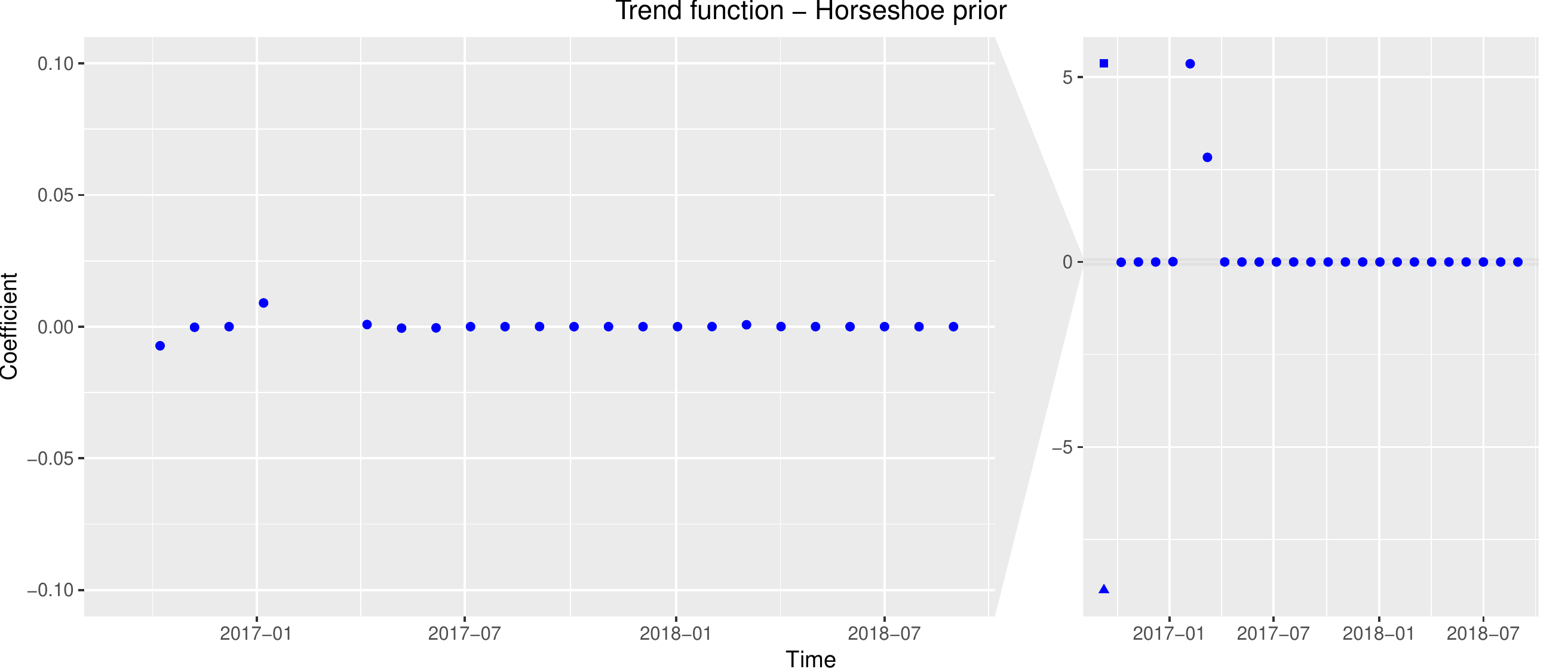}

  \caption{Trend coefficients of \negBinomModel on the time series B2.}
  \label{B2ModelTrendNegBinom}
\end{figure*}

In this section, the time series B2 (compare to Section \ref{ssec:data}) corresponding to \restB is considered.
This time series shows a drastic trend change within a short period of time.
In Figure \ref{B2ModelFitNegBinom}, the model fit of  \negBinomModel is
visualized with different prior specifications of the trend function.
We use the prior     \eqref{eq:priorNegBin}  --  \eqref{eq:normalPriorCoeff},
that is inspired by the Bayesian Lasso, and also the horseshoe prior.
In case of prior  \eqref{eq:priorNegBin}  --  \eqref{eq:normalPriorCoeff} the model is
trained once with the standard settings    \eqref{eq:standardSettings}
and once according to the
step-wise cross-validation (CV) described in Section \ref{overallStrategy}.
The cross-validation is performed using $6$ train/test splits with the testing sets including $14$ observations, respectively.
In  Figures $\ref{B2ModelFitNegBinom}$ and $\ref{B2ModelTrendNegBinom}$, we observe that the model
has some serious problems in case that prior
\eqref{eq:priorNegBin}  --  \eqref{eq:normalPriorCoeff} is applied with the
standard specification of the tuning parameters. The shrinkage of the trend
coefficients is too strong, such that the trend function fails to sufficiently
model the abrupt decrease at the end of 2017. If cross-validation is performed,
a better model fit is obtained. However, significant trend changes are detected
nearly every month which is an indication of overfitting. Due to the uniform
shrinkage going along with the Lasso prior, insignificant trend changes cannot
be shrunken sufficiently in order to allow for the appearance of drastic changes.
Finally, an application of the horseshoe prior results in a good model fit.
The trend function includes few large coefficients in absolute values to model
the abrupt trend change, while most of the coefficients are equal to zero.

\label{ssec:eval2}
\subsection{Comparison Against Other Prediction Methods}
In this section, the performance of our models is compared against other well-established
time series forecasting approaches. The priors given by the Equations  \eqref{eq:priorNormal} -- \eqref{eq:normalPriorCoeff}
are assigned to
the trend functions of our models and, further, the
standard settings specified
in \eqref{eq:standardSettings} for the model tuning parameters are used.

For a given time series and a given accuracy measure the performance of the
considered model is evaluated using a $15$-fold cross-validation as illustrated
in Figure \ref{fig:cv}. The size of the test datasets is specified as $14$.

\subsubsection{Forecast Accuracy Measures}
\label{sec:measures}
Clearly, some measures to compare the prediction quality of our
approach against others are required. Let $O=\{t_1,\ldots,t_n\}$
denote the set of training time points and, further, let $F=\{t_1^*,\ldots,t_p^*\}$
denote the set of testing time points. Additionally, we denote the observed
values with $y_t$ and the predicted ones with $\widehat{y}_t$.
A review of related work, see Section \ref{sec:relatedWork},
shows that commonly used measures for point estimates are \cite{LIU2017, Taylor2018}:
\begin{compactitem}
  \item \textit{mean absolute deviation}: $\text{MAD}=\frac{1}{p}\sum\limits_{t\in F}|y_t-\widehat{y}_t|$,
  \item \textit{mean squared error}: $\text{MSE}=\frac{1}{p}\sum\limits_{t\in F}(y_t-\widehat{y}_t)^2$,
  \item \textit{mean absolute percentage error}: $\text{MAPE}=\frac{1}{p}\sum\limits_{t\in F}\left|\frac{y_t-\widehat{y}_t}{y_t}\right|$.
\end{compactitem}
Another common measure is the \textit{mean forecast error}
\begin{align*}
  % \label{eq:MFE}
  \text{MFE} = \frac{1}{p}\sum\limits_{t\in F} (y_t-\widehat{y}_t),
\end{align*}
that is positive if a model tends to under-forecast and negative otherwise.
Except of MAPE all up to now mentioned measures are scale dependent, i.e.,
the larger the $y_t$ the larger the error.

We compare our approach against others
from the
literature on several different time series, i.e., different menu items and restaurants,
in order to obtain a broad comparison.
Hence, it is desirable that the used prediction quality measures are
scale invariant. The above mentioned scale dependent
measures are important because in the restaurant industry the calculation of the nominal and safety stocks are based on these measures. In case of inferior estimates one possible encounters inventory shortages or increased food waste.
For this reason, the scale dependent measures are considered in
Section \ref{sec:eval}, but with the note that the aggregated values presented
in this section are dominated by those time series that show the largest values on average.
The scale invariant MAPE is not considered since it can not deal with zero values,
i.e., $y_t$ must be different from zero.
Some more  suitable measures are the \textit{mean arctangent absolute percentage error} (MAAPE),
proposed by \citet{KIM2016669},
the  \textit{weighted absolute percentage error} (WAPE) \cite{Chase1995},
and the \textit{mean absolute scaled error} (MASE) \cite{HYNDMAN2006, MAKRIDAKIS2020}:
\begin{compactitem}
  \item $\text{MAAPE}=\frac{1}{p}\sum\limits_{t\in F}\operatorname{arctan}\left(\left|\frac{y_t-\widehat{y}_t}{y_t}\right|\right)$,
  \item $\text{WAPE}=\frac{\sum\limits_{t\in F}|y_t-\widehat{y}_t|}{\sum\limits_{t\in F}y_t}$,
  \item $\text{MASE}=\frac{\frac{1}{p}\sum\limits_{t\in F}|y_t-\widehat{y}_t|}{\operatorname{mean}(\operatorname{abs}(\mathbf{y}-\mathbf{y}_m))}$,
\end{compactitem}
with $m\in\mathbb{N}, ~\mathbf{y}=\transpose{(y_{t_1+m},\ldots,y_{t_n})}$ and $\mathbf{y}_m=\transpose{(y_{(t_1+m)-m},\ldots,y_{t_n-m})}$.
MAAPE can be considered  as a modification of MAPE that allows for
zero values since it passes the absolute percentage errors to the $\operatorname{arctan}$ function.
In contrast to MAPE, WAPE weights errors by the sales volume.
MASE compares the MAD of the considered forecasting model (in the testing period $F$)
with the MAD of the one-step seasonal naive method with season $m$ (in the training period $O$).
It should be mentioned that in Section \ref{sec:eval} MASE is once used with $m=1$ (MASE1) and once used with $m=7$ (MASE7).

In order to evaluate the quality of the prediction intervals the following measures are helpful:
\textit{prediction interval coverage probability} (PICP) \cite{Shen2018},
\textit{prediction interval normalized average width} (PINAW) \cite{Shen2018},
\textit{mean scaled interval score} (MSIS) \cite{Gneiting2007, MAKRIDAKIS2020}.
It is desirable to have a small PINAW and a large PICP \cite{Abbas2010}.
MSIS is designed to balance the relationship between PINAW and PICP.
Note that in this evaluation study always $95\%$ prediction intervals are considered,
since intervals with this confidence level are most commonly used in the business world \cite{MAKRIDAKIS2020}.

\subsubsection{Prediction Methods to Compare Against}

Now, we briefly present the methods that we consider for comparison against our proposed
model. Therefore, we adopt the selection taken by \citet{Taylor2018}.

\begin{compactitem}
  \item \textbf{Facebook's \texttt{prophet} method.}
  The \texttt{prophet} \R package implements a modular regression model with interpretable parameters \cite{Taylor2018}.
  We discuss the model in more detail in Section  \ref{sec:relatedWork}.
  Within the performance comparison the model is once used with automatic season
  detection (prophetA) and once with yearly and weekly seasonality (prophetS).
  \item
  \textbf{\ARIMA.}
  The \texttt{auto.arima} function returns the best \ARIMA model according
  to the AIC or BIC value. Here, we choose the Akaike information criterion (AIC).
  \item
  \textbf{Seasonal ARIMA.}
  The \texttt{auto.arima} function can also be used to determine the best seasonal ARIMA (SARIMA) model.
  To this aim, we specify the seasonality to be weekly and, further,
  define the order of seasonal differencing equal to $1$.
  If the seasonal order is not specified, the consideration of a seasonality
  is decided by the \texttt{auto.arima} function which might also return a non-seasonal model.

  \item
  \textbf{Exponential Smoothing.}
  The   \texttt{ets} function implements an exponential smoothing state space model.
  It fits a collection of exponential smoothing models and selects the best, see \citet{HYNDMAN2002439}.
  In this performance comparison the seasonality of the exponential smoothing model is defined to be weekly.

  \item \textbf{TBATS.}
  Trigonometric seasonality based on Fourier series, Box-Cox transformation, ARMA errors, Trend, and Seasonal components,
  see   \citet{DeLivera2011}. In this comparison study, the TBATS model is specified to learn weekly, monthly, and yearly seasonalities.
\end{compactitem}
Note that all above mentioned forecasting models, except for \texttt{prophet},
are implemented in the \texttt{forecast} \R package \cite{forecast2008, forecast2019}.
We want to point out that \ARIMA, SARIMA, ExpSmooth,
and TBATS assume regularly spaced data points. Since the considered testing data
(compare to Section \ref{sec:assumptions} and Section \ref{ssec:data}) contains
missing data, the missing values are linearly interpolated and subsequently
rounded before fitting one of these models.

\subsubsection{Evaluation}
\label{sec:eval}
Finally, we present how our approach performs compared to the ones described above.
In  Tables \ref{tab:evaluationRestA_starters} --
\ref{tab:evaluationRestB} averages of the performance measures described in Section \ref{sec:measures} are given,
respectively per dataset/category.
For each single time series the measures are computed based on a $15$-fold cross-validation.

Considering the results for \restA it can be seen that both of our models show a
positive MFE for nearly all of the considered categories.
This indicates, that our approach tends to under-forecast the sales numbers.
While also tending to under-forecasting, the classical times series models (ARIMA, SARIMA, ExpSmooth, TBATS)
show the lowest and, thus, the best MFEs.
However, none of the models under comparison shows large MFEs, such that one could speak about a poor performance.
In contrast to \restA, for \restB the non-standard approaches
(i.e.,\normalModel, \negBinomModel, and \texttt{prophet}) obtain lower MFEs than the classical ones.
Especially, our models show significantly lower values than all other methods under comparison.

Inspecting the MSEs and the MADs for \restA reveals that the proposed models often obtain the lowest errors.
In case that another model performs better than ours, the differences are rather small.
For \restB \negBinomModel obtains the lowest MAD and the second lowest MSE.

Switching the analysis from the scale-dependent metrics for point forecasts to the scale-invariant ones
(WAPE, MAAPE, MASE) shows that our approach \negBinomModel shows the best results overall.
For \restB \negBinomModel obtains significantly lower error values than all other considered approaches.
This result is stronger than the result for the scale-dependent metrics,
since the scale-invariant metrics are not dominated by the time series showing the highest number of sales.

Moreover, \negBinomModel always achieves
the lowest mean scaled interval score. Therefore, the prediction intervals corresponding to
this approach can be considered as the most useful ones.
The superior performance could be explained by a good coverage probability paired with tight interval widths.

Finally, it is important to link the performance of the methods to the structural characteristics of the time series data.
In Section \ref{ssec:data}, it is shown that the individual time series show multiple seasonalities.
While, weekly, monthly, and yearly seasonal effects are observed, the strength of each of these effects differs among menu items.
In principle, a method being capable of incorporating each of the mentioned seasonalities should return the best results.
However, in case of limited training data one has to be careful.
Allowing for too much flexibility, i.e., modeling each of the seasons, can lead to over-fitting.
This may be the reason why TBATS often shows inferior results than some of the simpler forecasting approaches.
On the other hand, considering only some of the seasonalities can also negatively impact the model performance.
For instance, the SARIMA model, that only considers weekly seasonalities,
most of the times returns worse point predictions than the simple ARIMA model.
We assume that neglecting the remaining seasonalities makes it difficult to
extract the true weekly effects, especially if the amount of training data is limited.
Thus, we conclude that an approach which considers all of the mentioned seasonal
effects, but at the same time, in terms of regularization, also penalizes the model flexibility should be preferred.
This observations agree with the fact that our approach leads to the best predictions overall.
\normalModel and \negBinomModel use shrinkage priors in
order to automatically detect significant seasonalities and, further,
to guarantee that their influence does not exceed a necessary level.

Finally, another reason for the superior performance is that all other methods somehow assume Gaussianity,
while the negative binomial distribution is more appropriate to model count data.
This might be the reason why \negBinomModel provides the most useful prediction intervals.
It is likely that the use of the negative binomial distribution is a key factor.
Similar findings are observed by \citet{snyder_forecasting_2012} who also
found that models based on the negative binomial distribution give the best
prediction performance concerning a forecast of the intermittent demand for
slow-moving inventories, i.e., low-volume items such as auto parts.
The negative binomial distribution is widely
used for slow-moving inventory items, presumably because of its good performance in practice.

\subsection{Evaluation of the Day of Month Effect}

While it is quite intuitive that sales of menu items show weekly
and yearly patterns, it is less evident that there are also effects that
repeat monthly.
As shown in Section \ref{ssec:data} the proposed models detect monthly
seasonalities within  the time series data considered in this study.
However, there is still the question if these effects significantly influence the model prediction accuracy.
  To answer this question, we have repeated the evaluation performed
in Section \ref{sec:eval} for \normalModel and \negBinomModel with the
\textit{day of month} effect switched off. It turns out that the usage of this explanatory
variable does not have a significant influence on the average model
performance, whether for \RestA or  \RestB.
However, a more detailed
analysis shows that there are some time series where the consideration
the \textit{the day of month}  effect leads to consistent improvements.
For instance, time series A2 (compare to Section \ref{ssec:data}) is
such a time series. In Table \ref{tab:evaluationRestA2},
the corresponding model accuracies are reported once with the
\textit{day of month} effect turned on and once with this effect turned off.
We notice that the consideration of monthly patterns leads
to superior accuracies.
Since the additional model
complexity introduced by
the \textit{day of month} effect has only a
% the usage of monthly effects has only a
minor influence on the computation times, we conclude that their
consideration is useful.
However, we found that the model performance
 can only be improved for a  fraction of the restaurant sales time series considered in this paper.

\setlength\tabcolsep{3pt}
\begin{table}[!ht]
  \scriptsize
  \centering
  % !TeX spellcheck = en_US
\begin{tabular}{
  l
  *{8}S[table-format=1.3]
  }
  \hline
  {Measure}  &    {Normal}   &    {NegBinom}    &    {ProphetS}   &    {ProphetA}     &    {ARIMA}     &    {SARIMA}    &    {ExpSmooth}   &    {TBATS}  \\
  \hline
  MFE & 0.3568735 & 0.5968023 & 0.2357142 & 1.266542 & 0.3757917 & 0.5429607 & 0.318535 & \maxf{-0.05947598} \\
  MSE & 14.49493 & 13.67789 & 13.86046 & 17.10028 & 14.16821 & 14.04731 & \maxf{12.9075} & 13.78354 \\
  MAD & 2.675002 & 2.512172 & 2.644101 & 2.853282 & 2.504553 & 2.532528 & \maxf{2.457947} & 2.584678 \\\\
  WAPE & 0.8407768 & \maxf{0.8108828} & 0.8135255 & 0.9950938 & 0.8747736 & 0.8677385 & 0.833637 & 0.8367019 \\
  MAAPE & 0.6609554 & \maxf{0.6522437} & 0.6672117 & 0.6698003 & 0.668004 & 0.6631046 & 0.6584866 & 0.6726629 \\
  MASE1 & 0.8156135 & 0.7508751 & 0.8137883 & 0.8655058 & 0.745824 & 0.7527794 & \maxf{0.7353459} & 0.7759299 \\
  MASE7 & 0.8199606 & 0.757583 & 0.8170813 & 0.8738278 & 0.754783 & 0.7633242 & \maxf{0.7451597} & 0.7828368 \\\\
  PICP & 0.9266667 & 0.9352381 & 0.9257143 & 0.9038095 & 0.9447619 & 0.9380952 & 0.9428571 & 0.9390476 \\
  PINAW & 2.243829 & 1.374707 & 2.231369 & 2.435371 & 2.536522 & 2.527794 & 2.475333 & 2.29007 \\
  MSIS1 & 6.436358 & \maxf{5.458995} & 6.388803 & 7.480836 & 6.542818 & 6.503957 & 6.060836 & 6.250627 \\
  MSIS7 & 6.478731 & \maxf{5.461642} & 6.406346 & 7.549797 & 6.615704 & 6.574803 & 6.122643 & 6.295626 \\
  \hline
\end{tabular}

  \caption{Performance comparison based on data from category \textit{starters} collected at \restA.}
  \label{tab:evaluationRestA_starters}
\end{table}

\begin{table}[!ht]
  \scriptsize
  \centering
  % !TeX spellcheck = en_US
\begin{tabular}{
  l
  *{8}S[table-format=2.3]
  }
  \hline
  {Measure}  &    {Normal}   &    {NegBinom}    &    {ProphetS}   &    {ProphetA}     &    {ARIMA}     &    {SARIMA}    &    {ExpSmooth}   &    {TBATS}  \\
  \hline
  MFE & -0.1702228 & -0.1322703 & \maxf{-0.1133266} & -0.188105 & -0.1262269 & -0.2342264 & -0.149586 & -0.1638112 \\
  MSE & 8.227617 & 8.164722 & \maxf{8.070301} & 8.368138 & 8.17702 & 9.778167 & 8.195514 & 8.475017 \\
  MAD & 2.073194 & 2.064481 & 2.066681 & 2.08812 & 2.058756 & 2.249842 & \maxf{2.053722} & 2.127686 \\\\
  WAPE & 0.6316235 & \maxf{0.6279537} & 0.6291264 & 0.655165 & 0.6326891 & 0.6886623 & 0.6308128 & 0.6548442 \\
  MAAPE & 0.598347 & 0.5972028 & 0.5962342 & 0.6025624 & 0.5983719 & 0.6196971 & \maxf{0.5939586} & 0.6104913 \\
  MASE1 & 0.7679289 & \maxf{0.7656286} & 0.7681626 & 0.7833588 & 0.7719206 & 0.8370499 & 0.7694234 & 0.7992026 \\
  MASE7 & 0.7479293 & \maxf{0.7454851} & 0.7482035 & 0.7608705 & 0.750405 & 0.8140961 & 0.7478966 & 0.7762681 \\\\
  PICP & 0.9428571 & 0.9659864 & 0.9455782 & 0.9482993 & 0.9537415 & 0.9564626 & 0.9517007 & 0.937415 \\
  PINAW & 1.329892 & 1.247955 & 1.373262 & 1.434302 & 1.471097 & 1.652499 & 1.459302 & 1.389725 \\
  MSIS1 & 5.489743 & \maxf{4.792476} & 5.343381 & 5.353174 & 5.414221 & 5.604436 & 5.324117 & 5.42939 \\
  MSIS7 & 5.317548 & \maxf{4.646741} & 5.173699 & 5.198237 & 5.251483 & 5.431977 & 5.163297 & 5.252192 \\
  \hline
\end{tabular}

  \caption{Performance comparison based on data from category \textit{side dishes} collected at \restA.}
  \label{tab:evaluationRestA_sides}
\end{table}

\begin{table}[!ht]
  \scriptsize
  \centering
  % !TeX spellcheck = en_US
\begin{tabular}{
  l
  *{8}S[table-format=2.3]
  }
  \hline
  {Measure}  &    {Normal}   &    {NegBinom}    &    {ProphetS}   &    {ProphetA}     &    {ARIMA}     &    {SARIMA}    &    {ExpSmooth}   &    {TBATS}  \\
  \hline
  MFE & 0.2549779 & 0.2085412 & 0.1575834 & 0.3357646 & 0.1102064 & 0.1203235 & \maxf{0.05787903} & 0.06518352  \\
  MSE & 14.13285 & \maxf{13.97443} & 14.00075 & 15.67997 & 15.53779 & 16.89321 & 14.92484 & 15.28157  \\
  MAD & 2.680936 & \maxf{2.675391} & 2.693156 & 2.799868 & 2.822731 & 2.96366 & 2.754525 & 2.810389  \\\\
  WAPE & 0.5067425 & \maxf{0.5055266} & 0.5160045 & 0.5256489 & 0.5326327 & 0.5669093 & 0.51776 & 0.5420104 \\
  MAAPE & \maxf{0.5086675} & 0.5097708 & 0.5107298 & 0.5147795 & 0.5194417 & 0.539119 & 0.5133304 & 0.5243478  \\
  MASE1 & 0.7790618 & \maxf{0.7774146} & 0.7886293 & 0.8112173 & 0.8179948 & 0.8661218 & 0.7981058 & 0.8293299  \\
  MASE7 & 0.7851178 & \maxf{0.7832212} & 0.794164 & 0.8173644 & 0.8266428 & 0.8722089 & 0.8046634 & 0.834852  \\\\
  PICP & 0.9281179 & 0.9530612 & 0.9276644 & 0.9326531 & 0.9387755 & 0.9482993 & 0.937415 & 0.9160998  \\
  PINAW & 1.211489 & 1.168017 & 1.243072 & 1.308402 & 1.368231 & 1.550877 & 1.349633 & 1.251989  \\
  MSIS1 & 5.638205 & \maxf{4.761095} & 5.666142 & 5.867706 & 5.810629 & 5.786575 & 5.571586 & 6.236446  \\
  MSIS7 & 5.695481 & \maxf{4.798255} & 5.706415 & 5.927439 & 5.876799 & 5.844616 & 5.629913 & 6.260858  \\
  \hline
\end{tabular}

  \caption{Performance comparison based on data from category \textit{main dishes} collected at \restA.}
  \label{tab:evaluationRestA_main}
\end{table}

\begin{table}[!ht]
  \scriptsize
  \centering
  % !TeX spellcheck = en_US
\begin{tabular}{
  l
  *{8}S[table-format=1.3]
  }
  \hline
  {Measure}  &    {Normal}   &    {NegBinom}    &    {ProphetS}   &    {ProphetA}     &    {ARIMA}     &    {SARIMA}    &    {ExpSmooth}   &    {TBATS}  \\
  \hline
  MFE & 0.3017014 & 0.1726079 & 0.1676618 & 0.4499957 & 0.06023987 & 0.08368891 & 0.101739 & \maxf{0.006686447} \\
  MSE & 5.845566 & 5.390144 & 5.79594 & 5.748678 & \maxf{5.383916} & 5.794933 & 5.461953 & 5.929645 \\
  MAD & 1.459947 & \maxf{1.453194} & 1.513747 & 1.469039 & 1.481701 & 1.543236 & 1.482757 & 1.576512 \\\\
  WAPE & 0.8625233 & 0.8632097 & 0.882095 & \maxf{0.8444306} & 0.880569 & 0.9037678 & 0.8618778 & 0.9566893 \\
  MAAPE & \maxf{0.76613} & 0.7706784 & 0.7890387 & 0.7853289 & 0.782571 & 0.8096771 & 0.7816391 & 0.8058513 \\
  MASE1 & 0.7253126 & \maxf{0.7238416} & 0.7455421 & 0.7267039 & 0.7379614 & 0.7673786 & 0.7303072 & 0.7854205 \\
  MASE7 & 0.7369345 & \maxf{0.735411} & 0.7572112 & 0.7386781 & 0.7499737 & 0.7791245 & 0.7421004 & 0.7988843 \\\\
  PICP & 0.9530612 & 0.9680272 & 0.9503401 & 0.9557823 & 0.9639456 & 0.9680272 & 0.9646259 & 0.9571429 \\
  PINAW & 1.604493 & 1.233461 & 1.652187 & 1.714701 & 1.76403 & 1.957152 & 1.75682 & 1.663627 \\
  MSIS1 & 5.690034 & \maxf{4.539035} & 5.586051 & 5.975866 & 5.781902 & 5.966652 & 5.720662 & 5.75078 \\
  MSIS7 & 5.767792 & \maxf{4.610346} & 5.663471 & 6.05545 & 5.861913 & 6.047259 & 5.798664 & 5.821798 \\
  \hline
\end{tabular}

  \caption{Performance comparison based on data from category \textit{desserts} collected at \restA.}
  \label{tab:evaluationRestA_dessert}
\end{table}

\begin{table}[!ht]
  \scriptsize
  \centering
  % !TeX spellcheck = en_US
\begin{tabular}{
  l
  S[table-format=2.3]
  S[table-format=2.3]
  S[table-format=2.3]
  S[table-format=2.3]
  S[table-format=2.3]
  S[table-format=2.3]
  S[table-format=2.3]
  S[table-format=2.3]
  }
  \hline
  {Measure}  &    {Normal}   &    {NegBinom}    &    {ProphetS}   &    {ProphetA}     &    {ARIMA}     &    {SARIMA}    &    {ExpSmooth}   &    {TBATS}  \\
  \hline
  MFE & 0.1459185 & 0.1074649 & 0.1024824 & 0.209968 & \maxf{-0.001589677} & 0.1014397 & -0.002764375 & -0.081451  \\
  MSE & 12.31975 & 12.18444 & \maxf{12.04435} & 14.03706 & 12.70386 & 16.11357 & 12.90098 & 13.92804  \\
  MAD & \maxf{1.919412} & 1.923863 & 1.948215 & 1.986885 & 1.979871 & 2.229552 & 1.9832 & 2.054397 \\\\
  WAPE & \maxf{0.6113923} & 0.6127004 & 0.6223718 & 0.6225019 & 0.6329124 & 0.7136776 & 0.630206 & 0.6443825 \\
  MAAPE & \maxf{0.5864744} & 0.5874837 & 0.5945489 & 0.5911441 & 0.5954454 & 0.6347464 & 0.5958744 & 0.6048865  \\
  MASE1 & \maxf{0.7496775} & \maxf{0.7503199} & 0.7635177 & 0.7679587 & 0.7735695 & 0.8732264 & 0.7724184 & 0.7908049  \\
  MASE7 & \maxf{0.7465738} & \maxf{0.7471562} & 0.7605147 & 0.7644592 & 0.7705509 & 0.8702813 & 0.768815 & 0.7868759  \\\\
  PICP & 0.9407738 & 0.9672619 & 0.9377976 & 0.9452381 & 0.9532738 & 0.9565476 & 0.9505952 & 0.9389881  \\
  PINAW & 1.227338 & 1.161172 & 1.265572 & 1.321901 & 1.373465 & 1.641748 & 1.350288 & 1.278358  \\
  MSIS1 & 5.24769 & \maxf{4.496459} & 5.207521 & 5.47171 & 5.317956 & 5.975563 & 5.257222 & 5.323224  \\
  MSIS7 & 5.206548 & \maxf{4.463839} & 5.175618 & 5.427682 & 5.284129 & 5.941989 & 5.219886 & 5.284743  \\
  \hline
\end{tabular}

  \caption{Performance comparison based on data from category \textit{snacks} collected at \restA.}
\end{table}

\begin{table}[!ht]
  \scriptsize
  \centering
  % !TeX spellcheck = en_US
\begin{tabular}{
  l
  S[table-format=2.3]
  S[table-format=2.3]
  S[table-format=2.3]
  S[table-format=3.3]
  S[table-format=3.3]
  S[table-format=3.3]
  S[table-format=2.3]
  S[table-format=7.3]
  }
  \hline
  {Measure}  &    {Normal}   &    {NegBinom}    &    {ProphetS}   &    {ProphetA}     &    {ARIMA}     &    {SARIMA}    &    {ExpSmooth}   &    {TBATS}  \\
  \hline
  MFE & 0.3926242 & 0.3900117 & 0.3242011 & 0.5671003 & -0.03443436 & \maxf{0.00214872} & 0.0627411 & 27.11415 \\
  MSE & \maxf{84.78169} & 85.73686 & 86.01201 & 102.7643 & 122.2761 & 101.1332 & 97.73978 & 1124662.000 \\
  MAD & \maxf{4.518726} & 4.525069 & 4.657049 & 4.788381 & 5.081013 & 5.086446 & 4.680459 & 40.09606 \\\\
  WAPE & 0.738744 & \maxf{0.7334583} & 0.8123215 & 0.750129 & 0.7628447 & 0.82397 & 0.7451321 & 12.04772 \\
  MAAPE & 0.6330027 & \maxf{0.6290948} & 0.6461365 & 0.6402016 & 0.6502532 & 0.6800015 & 0.6384264 & 0.658268 \\
  MASE1 & 0.7423235 & \maxf{0.7398369} & 0.8071795 & 0.765024 & 0.7826782 & 0.8584485 & 0.7560752 & 15.88204 \\
  MASE7 & 0.7442101 & \maxf{0.7416765} & 0.8130403 & 0.7674875 & 0.7860833 & 0.8608813 & 0.7582058 & 18.41693 \\\\
  PICP & 0.9417989 & 0.9578483 & 0.9365079 & 0.943739 & 0.9502646 & 0.9546737 & 0.9518519 & 0.9292769 \\
  PINAW & 1.406165 & 1.226277 & 1.445418 & 1.523252 & 1.59337 & 1.846701 & 1.558657 & 1.458473 \\
  MSIS1 & 5.972293 & \maxf{4.947007} & 7.494552 & 6.200472 & 6.128499 & 6.541344 & 5.964309 & 609.6048 \\
  MSIS7 & 5.980117 & \maxf{4.95601} & 7.623279 & 6.216109 & 6.145332 & 6.55281 & 5.97491 & 710.8797 \\
  \hline
\end{tabular}

  \caption{Performance comparison based on data from category \textit{alcoholic beverages} collected at \restA.}
  \label{tab:evaluationRestA_alc}
\end{table}

\begin{table}[!ht]
  \scriptsize
  \centering
  % !TeX spellcheck = en_US
\begin{tabular}{
  l
  S[table-format=1.3]
  S[table-format=1.3]
  S[table-format=1.3]
  S[table-format=1.3]
  S[table-format=1.3]
  S[table-format=2.3]
  S[table-format=1.3]
  S[table-format=1.3]
  }
  \hline
  {Measure}  &    {Normal}   &    {NegBinom}    &    {ProphetS}   &    {ProphetA}     &    {ARIMA}     &    {SARIMA}    &    {ExpSmooth}   &    {TBATS}  \\
  \hline
  MFE & 0.1672703 & 0.1276753 & 0.1127017 & 0.1087032 & 0.03313573 & 0.02731743 & \maxf{0.007350705} & 0.07676429 \\
  MSE & 8.63913 & \maxf{8.603087} & 8.931199 & 9.345957 & 9.309124 & 10.80058 & 9.241184 & 9.746363 \\
  MAD & 2.017433 & \maxf{2.013264} & 2.063239 & 2.079467 & 2.091063 & 2.241962 & 2.08128 & 2.134773 \\\\
  WAPE & 0.7556816 & \maxf{0.7539938} & 0.7741326 & 0.7787194 & 0.7716322 & 0.8342483 & 0.7635799 & 0.7841502 \\
  MAAPE & \maxf{0.6837305} & \maxf{0.6835529} & 0.6992984 & 0.6910114 & 0.6948896 & 0.7277045 & 0.6940355 & 0.7060461 \\
  MASE1 & 0.7618515 & \maxf{0.7603541} & 0.7799606 & 0.7809311 & 0.7848934 & 0.8456421 & 0.7797006 & 0.7998393 \\
  MASE7 & 0.7678029 & \maxf{0.7662563} & 0.7858194 & 0.7869345 & 0.7909958 & 0.8521367 & 0.7854077 & 0.8061144 \\\\
  PICP & 0.9403274 & 0.9672619 & 0.9370536 & 0.9473214 & 0.9497024 & 0.9517857 & 0.9495536 & 0.9348214 \\
  PINAW & 1.414125 & 1.227556 & 1.457502 & 1.528093 & 1.556707 & 1.792123 & 1.555826 & 1.485368 \\
  MSIS1 & 5.724734 & \maxf{4.522523} & 5.716661 & 5.924839 & 5.837592 & 6.210902 & 5.765902 & 5.969027 \\
  MSIS7 & 5.769615 & \maxf{4.558809} & 5.764296 & 5.971963 & 5.886676 & 6.259513 & 5.81179 & 6.019709 \\
  \hline
\end{tabular}

  \caption{Performance comparison based on data from category \textit{non-alcoholic beverages} collected at \restA.}
  \label{tab:evaluationRestA_nonalc}
\end{table}

\begin{table}[!ht]
  \scriptsize
  \centering
  % !TeX spellcheck = en_US

\begin{tabular}{
  l
  *{8}S[table-format=5.3]
  }
  \hline
  {Measure}  &    {Normal}   &    {NegBinom}    &    {ProphetS}   &    {ProphetA}     &    {ARIMA}     &    {SARIMA}    &    {ExpSmooth}   &    {TBATS}  \\
  \hline
  MFE & -0.7348873 & \maxf{0.4809695} & -2.863709 & -3.589308 & -5.061448 & -9.297509 & -5.19148 & -7.497736 \\
  MSE & 37364.71 & 36975.5 & 40162.94 & 37331.88 & \maxf{36956.45} & 49030.34 & 38858.29 & 39497.44 \\
  MAD & 44.03961 & \maxf{41.1646} & 47.0454 & 44.48996 & 45.17681 & 49.54553 & 45.31106 & 46.75794 \\\\
  WAPE 			   &  0.9031184  &  \maxf{0.7363994}  &  0.9537209  &  0.8159774  &  0.8104929  &  0.9373068  &  0.837517  &  0.8547491 \\
  MAAPE & 0.6141947 &  \maxf{0.5800982} & 0.6405877 & 0.6017391 & 0.6149769 & 0.5882702 & 0.6208011 & 0.6210931 \\
  MASE1 & 0.923871 & \maxf{0.8395213} & 0.9854153 & 0.8396699 & 0.8571649 & 0.8994043 & 0.8899705 & 0.8939375 \\
  MASE7 & 0.9377283 &\maxf{0.8466499} & 0.9988078 & 0.8513444 & 0.8708477 & 0.9160105 & 0.9001735 & 0.9051161		 \\\\
  PICP & 0.9126984 & 0.91 & 0.8936508 & 0.9430159 & 0.9406349 & 0.9392063 & 0.9479365 & 0.914127 \\
  PINAW & 2.422543 & 1.653406 & 2.513839 & 2.657171 & 2.584516 & 2.964648 & 98.26026 & 2.360085 \\
  MSIS1 & 8.287771 & \maxf{6.61974} & 9.709142 & 7.497961 & 7.540283 & 7.825233 & 404.4657 & 8.099009 \\
  MSIS7 & 8.497117 & \maxf{6.793278} & 9.923391 & 7.677196 & 7.74966 & 8.051975 & 438.2527 & 8.23803 \\
  \hline

\end{tabular}

  \caption{Performance comparison based on data collected at \restB.}
  \label{tab:evaluationRestB}
\end{table}

\begin{table}[!ht]
  \scriptsize
  \centering
  % !TeX spellcheck = en_US
\begin{tabular}{
  l
  *{8}S[table-format=1.3]
  }
  \hline
  {Measure}  &    {Normal without monthly effect}   &    {Normal with monthly effect}    &    {NegBinom without monthly effect}   &    {NegBinom with monthly effect}  \\
  \hline
  MFE &-0.003631506 &\maxf{0.001658735}    &-0.01119195 & -0.01175246 \\
  MSE &4.057998     &3.986821       &4.006592    &\maxf{3.889596}  \\
  MAD &1.608995     &1.594948       &1.60126     &\maxf{1.580636}  \\\\
  WAPE &0.5962571   &0.591144       &0.5938315   &\maxf{0.5861061}  \\
  MAAPE &0.5926996  &0.5898213      &0.5916436   &\maxf{0.5876776}  \\
  MASE1 &0.7042057  &0.6980449      &0.700886    &\maxf{0.6918091}  \\
  MASE7 &0.7415646  &0.7350967      &0.7379993   &\maxf{0.7284697}  \\\\
  PICP &0.9571429   &0.95           &0.9690476   &0.9690476  \\
  PINAW &1.260362   &1.220559       &1.133475    &1.108623  \\
  MSIS1 &4.526777   &4.429215       &3.812071    &\maxf{3.740714} \\
  MSIS7 &4.763454   &4.659933       &4.011587    &\maxf{3.936838}  \\
  \hline
\end{tabular}

  \caption{Performance evaluation of the \textit{day of month} effect for time series A2.}
  \label{tab:evaluationRestA2}
\end{table}

\clearpage

\section{Conclusion \& Future Work}
\label{sec:conclusion}

In this work, a new approach for predicting future sales of menu items in restaurants
and staff canteens was proposed. In particular, two Bayesian generalized additive
models were presented. The first one assumes future sales to be normally distributed,
while the second one uses the more appropriate negative Binomial distribution.
Both approaches use shrinkage priors to automatically learn significant multiple
seasonal effects and trend changes. The features learned by the models have a
straightforward human interpretation which helps potential users to create the
necessary trust and confidence in the methodology. The performance of our approach
was extensively evaluated and compared to other well-established forecasting methods.
Basis of the analysis were two data sets retrieved from (electronic) Point of Sales (\POS) systems
collected at a restaurant and a staff canteen. The evaluations show that our
approach often provides the best and most robust point predictions overall.
Moreover, the prediction intervals of our model with the negative Binomial
distribution are notably more accurate than the ones obtained
by the other methods.

Currently, our approach only takes account of \POS data. This makes it
universally applicable, but also limits the prediction quality that can be achieved.
In future work, we plan to enrich the data source by weather data, information regarding
special events and holidays, and, last but not least, expert knowledge of restaurant managers.
Additionally, we want to extend our approach such that it can also predict on hourly basis.
Accurate sales predictions on hourly basis will allow an optimization of workforce planning.

\end{document}